\documentclass[a4paper,11pt]{article}
\usepackage{graphicx,amsmath,amsfonts,amssymb,dcolumn,amsthm,euscript}
\def\bea{\begin{eqnarray}}
\def\eea{\end{eqnarray}}
\def\be{\begin{equation}}
\def\ee{\end{equation}}
\usepackage[usenames,dvipsnames]{xcolor}
\usepackage{float}   

\usepackage{mathtools}
\DeclarePairedDelimiter\bra{\langle}{\rvert}
\DeclarePairedDelimiter\ket{\lvert}{\rangle}
\DeclarePairedDelimiterX\braket[2]{\langle}{\rangle}
{#1\,\delimsize\vert\,\mathopen{}#2}

\newcommand\nn{\nonumber} 
\newcommand{\bq}{\begin{equation}}
\newcommand\eq{\end{equation}}
\newcommand\pa{\partial}

\usepackage{changepage}

\usepackage{graphicx}

\newcommand\Ph{\varPhi}
\newcommand\Ps{\varPsi}

\usepackage{slashed}
\usepackage{stmaryrd}
\usepackage{verbatim}
\usepackage{blindtext}
\usepackage{amsmath}
\usepackage{geometry}
\usepackage{pdflscape}   
\usepackage[numbers,sort&compress]{natbib}
\graphicspath{{./images/}}
\usepackage{mathtools}
\usepackage{scrextend}  
\usepackage{jheppub}
\usepackage{cancel} 
\usepackage{bibentry}

\DeclareMathOperator{\Tr}{Tr}

\usepackage{booktabs}
\usepackage{caption}
\usepackage{float}
\usepackage{titlesec}
\usepackage{capt-of}

\usepackage{tcolorbox}
\usepackage{tabularx}
\usepackage{colortbl}
\tcbuselibrary{skins}

\newcolumntype{Y}{>{\raggedleft\arraybackslash}X}

\tcbset{tab/.style={enhanced,fonttitle=\bfseries,fontupper=\normalsize\sffamily,
colback=white,colframe=black,colbacktitle=white,
coltitle=black,center title}}

\usepackage{array}
\usepackage{arydshln}
\usepackage{mathtools}

\usepackage{hyperref}

\numberwithin{equation}{section}

\newcommand*{\transp}[2][-3mu] 
{\ensuremath{\mskip1mu\prescript{\smash{\mathrm t\mkern#1}}{}{\mathstrut#2}}} 

\newcommand{\overbar}[1]  
{\mkern 1.5mu\overline{\mkern-1.5mu#1\mkern-1.5mu}\mkern 1.5mu}

\theoremstyle{plain} 
\newtheorem{theorem}{Theorem}[section] 

\def\demi{\frac{1}{2}}

\def\Lie{\mathcal{L}}

\def\={&=&}

\def\bea{\begin{eqnarray}}
\def\eea{\end{eqnarray}}

\def\w{\wedge}
\def\p{\pa_z}
\def\p0{\pa_0}

\def\o {\omega}
\def\O {\varOmega}
\def\vp{\varphi}

\def\E0{\mathcal{E}^0  }

\def\Lie{\mathcal{L}}

\def\pbM{\begin{pmatrix}}
\def\peM{\end{pmatrix}}
\def\bM{\begin{matrix}}
\def\eM{\end{matrix}}

\def\C{\mathsf{C}}

\def\J{J_{\rm BRST}}

\def\fiu{f^{\mu i}}

\def\pu{\pa_\mu}
\def\pv{\pa_\nu}

\def\eps{\epsilon}

\title{BRST Noether Theorem and Corner~Charge~Bracket}

\author[a]{Laurent Baulieu,}
\author[a]{Tom Wetzstein,}
\author[b]{and Siye Wu}

\affiliation[a]{LPTHE, Sorbonne Universit\'e, CNRS,
4 Place Jussieu, 75005 Paris, France}
\affiliation[b]{Department of Mathematics, National Tsing Hua University,
Hsinchu 30013, Taiwan}

\emailAdd{baulieu@lpthe.jussieu.fr}
\emailAdd{twetzstein@lpthe.jussieu.fr}
\emailAdd{swu@math.nthu.edu.tw}

\abstract{We provide a proof of the BRST Noether 1.5th theorem, conjectured
in [\href{https://doi.org/10.1007/JHEP10(2024)055}{\textit{JHEP} \textbf{10}
(2024) 055}], for a broad class of rank-$1$ BV theories including supergravity
and $2$-form gauge theories.
The theorem asserts that the BRST Noether current of any BRST invariant gauge
fixed Lagrangian decomposes on-shell into a sum of a BRST-exact term and a
corner term that defines Noether charges.
This extends the holographic consequences of Noether’s second theorem to
gauge fixed theories and, in particular, offers a universal gauge independent
Lagrangian derivation of the invariance of the $\mathcal{S}$-matrix under
asymptotic symmetries. 
Furthermore, we show that these corner Noether charges are inherently
non-integrable.   
To address this non-integrability, we introduce a novel charge bracket that
accounts for potential symplectic flux and anomalies, providing an honest
canonical representation of the asymptotic symmetry algebra.
We also highlight a general origin of a BRST cocycle associated with
asymptotic symmetries.}

\begin{document} 
\maketitle

\date{$ $\today}

\section{Introduction} 

More than a century ago, Emmy Noether discovered that classical gauge theories
admit an infinite number of codimension-two quantities $q_\lambda$, labeled by
the local gauge parameter~$\lambda(x)$~\cite{Noether:1918zz}. 
Since then, it has been understood that these quantities are the conserved
charges associated with the gauge symmetry, supported on the corners of the
Lorentzian spacetime manifold~$M$. 
This result, known as \textit{Noether's second theorem}, has profound
implications for holography, which have been extensively studied over the
last decade (see~\cite{Ciambelli:2022vot} for an overview of the literature). 


Notably, the Noether charges $q_\lambda$ associated with the invariance of
a Lagrangian $L(\phi,\pa_\mu\phi)$ under gauge transformations
$\delta_\lambda\phi=f(\phi)\lambda+f^\mu(\phi)\pa_\mu\lambda$ play a crucial
role in defining asymptotic symmetries
\cite{bondi,sachs,Brown:1986nw,BMS/CFT,Strominger_2014,He:2014cra}.
The \textit{asymptotic symmetries} or, in other words, the 
\textit{large gauge symmetries}\footnote{Be aware that,
even if they share the same terminology, such gauge transformations have
nothing to do with transformations that are not continuously connected to
the identity.}
are defined as the gauge transformations parametrized by $\lambda(x)$ that
preserve both the falloffs of the fields near the boundary
$\varSigma=\pa M$ and the gauge fixing condition, and that lead to a
non-vanishing Noether charge
\begin{equation}\label{Noether_corner_charge_classical}
Q_\lambda\equiv\int_{\pa\varSigma}q_\lambda\ne0 .
\end{equation} 
These large gauge symmetries generate non-trivial physical transformations on
the phase space of the theory and are no longer mere gauge redundancies.  

With the discovery of the infrared triangle
\cite{Strominger_lectures,BMS_soft_graviton,BMS_memory,campiglia2015asymptotic},
it is now understood that asymptotic symmetries govern the infrared behavior
of a given gauge theory, as they are equivalent to the soft theorem of the
associated massless gauge particle
\cite{Kapec,Campiglia,BMS_soft_graviton,campiglia2015asymptotic,BMS_memory}.
To establish this equivalence, one must assume the Hamiltonian Ward identity  
\begin{equation}\label{strominger_claim}
\bra{\rm out}[Q_\lambda,\mathcal{S}]\ket{\rm in}=0,
\end{equation}
which expresses the invariance of the physical $\mathcal{S}$-matrix under
asymptotic symmetries \cite{Strominger_BMS_scattering}.

While this assumption has been widely employed, asymptotic symmetries \eqref{Noether_corner_charge_classical} are defined at the classical level for a given choice of gauge; thus, a proper BRST-invariant gauge-fixing procedure is necessary to rigorously justify \eqref{strominger_claim} at the quantum level and to demonstrate its gauge independence.

This Lagrangian BRST analysis was carried out in \cite{Baulieu:2024oql} to clarify this ambiguity,  but only for the cases of Yang--Mills theory and gravity,  by explicitly computing the Noether current associated with the residual BRST symmetry in specific gauges, each chosen for its distinct physical relevance.
In all of these cases, the BRST Noether current could be expressed as
\begin{equation}
\label{conjecture_Noether_1,5}
J^\mu_{\rm BRST}\ \hat{=}\ s G^\mu_{\rm gauge}
+\pa_\nu\big(q^{\mu\nu}_{\text{cl}} + q^{\mu\nu}_{\rm gauge} \big),
\end{equation}
where the symbol $\hat{=}$ denotes equality modulo the equations of motion of the chosen gauge fixed and BRST-invariant action.\footnote{More specifically, in \cite{Baulieu:2024oql}, eq.~\eqref{conjecture_Noether_1,5} was proven for four-dimensional massless and massive Yang--Mills theory in all renormalizable gauges, as well as for four-dimensional gravity in the Bondi gauge. In these cases, it was found that $q_{\rm gauge}\ \hat{=}\ 0$, and the quantum Ward identity satisfied by the BRST current \eqref{conjecture_Noether_1,5} was provided, thereby demonstrating \eqref{strominger_claim} and its gauge independence at the perturbative quantum level.}

This striking expression, which encodes the cohomological structure of $J^\mu_{\rm BRST}$ with respect to $d$ and $s$, led the authors of \cite{Baulieu:2024oql} to propose that this property could be formulated as a general ``BRST Noether 1.5th theorem".

This denomination  $1.5$ was introduced because it generalizes  the second Noether theorem,  by using the fundamental property that the BRST symmetry is a global  (graded) symmetry. 
  It  allows one  to derive     \eqref{conjecture_Noether_1,5} directly from   the first Noether theorem,  within the context of the BRST symmetry,  while   trivially  including  the second theorem in the special case of an ungauge fixed action. 

Here the ghost number one corner charges $q_{\text{cl}}$ are the
classical Noether charges $q_\lambda$ in~\eqref{Noether_corner_charge_classical} when $\lambda$ is replaced by its corresponding ghost field $c$. The quantities $G_{\rm gauge}$ and $q_{\rm gauge}$ explicitly depend
on the choice of gauge.

There is little doubt about the validity of \eqref{conjecture_Noether_1,5}
for physically healthy theories, namely gauge anomaly free theories. 
Such theories obviously admit an LSZ formulated perturbative physical $\mathcal{S}$-matrix,  and soft theorems allow one to construct infrared safe observables \cite{Kulish:1970ut}. That the BRST Ward identity for the current \eqref{conjecture_Noether_1,5} is central to proving the soft theorems through \eqref{strominger_claim} then convinced us that \eqref{conjecture_Noether_1,5} must be generally correct.

This paper therefore aims to prove the validity of \eqref{conjecture_Noether_1,5} for a broad class of rank-1 BV theories and for any gauge-fixing functionals.

Another important feature of the Noether charges
\eqref{Noether_corner_charge_classical} is that their charge algebra  
projectively represents the asymptotic symmetry
algebra \cite{Brown:1986nw}. 
These charges canonically generate the action of asymptotic symmetries on
phase space.  
For asymptotically flat gravity, the asymptotic symmetry group is the BMS
group \cite{bondi,sachs} (and all its extensions
\cite{BMS/CFT,Campiglia,Compere,Freidel1,Geiller}) and the Noether charges
are non-integrable.  
Building a well-defined bracket for these charges that correctly represents
the BMS algebra and generates the right BMS transformations on phase space
has thus been a longstanding challenge \cite{Barnich_Charge_algebra,
Distler:2018rwu,Campiglia:2020qvc,Compere:2020lrt,Freidel1,Freidel,
Campiglia:2021bap,Donnay:2021wrk,Rignon-Bret:2024gcx,Geiller:2024amx,
Campiglia:2024uqq}.  
Since all these constructions have relied on Noether's second theorem, one
wonders whether an alternative construction exists for the BRST version of
Noether's theorem \eqref{conjecture_Noether_1,5}.

By analyzing the symplectic structure derived from the variation of any gauge fixed Lagrangian, we propose a new model independent bracket for
the non-integrable BRST Noether charges $q_{\text{cl}}$.
This bracket provides a centerless representation of the asymptotic symmetry algebra of the rank-$1$ BV theories considered here, for any gauge-fixing functional and remains valid in the presence of non-vanishing symplectic flux and symplectic anomaly.

The BRST symplectic structure also allows us to identify a model independent
ghost number two and spacetime codimension-two BRST cocycle for asymptotic
symmetries, generalizing the constructions in
\cite{Barnich_BRST,Baulieu_Tom_BMS}.
On-shell, this cocycle coincides with the Barnich--Troessaert bracket
\cite{Barnich_Charge_algebra}.  
This observation raises the possibility of finding a general ghost number one
and spacetime codimension-one BRST cocycle for asymptotic symmetries, which
would be responsible for loop corrections to soft theorems
\cite{Baulieu:2024oql}. 

The rest of the paper is organized as follows:   
\vspace{-0.4cm}\\
\begin{adjustwidth}{0.5em}{0cm} 
- In Section \ref{Section_BRST_Noether_current}, we define the rank-$1$ BV
theories under consideration and introduce the necessary tools for the
remainder of the paper.
We then provide a  detailed proof of the BRST 1.5th theorem and explore its
implications for holography. 
\vspace{0.2cm}\\
- In Section \ref{Section_symplectic}, we analyze the symplectic structure
arising from the symplectic potential of a BRST invariant gauge fixed
Lagrangian.  
We prove other conjectures in \cite{Baulieu:2024oql} concerning the gauge
dependence of the non-integrable part of the fundamental canonical relation.
This allows us to define a charge bracket that canonically represents the
asymptotic symmetry algebra of the theories under study for any gauge fixing
functional. 
\vspace{0.2cm}\\
- In Section \ref{Section_examples}, we apply the general results of the
previous sections to the theory of an abelian $2$-form coupled to
Chern--Simons.
We discuss the role of the non-vanishing gauge charge $q_{\rm gauge}$ that
appears in this case. 
\vspace{0.2cm}\\
- In Section \ref{Section_extension_rank-$2$}, we check the validity of the
BRST Noether theorem \eqref{conjecture_Noether_1,5} for an elementary rank-$2$
BV system, that is, for Yang-Mills theory in the Feynman-'t~Hooft gauge without the
$b$ field.
This suggests that \eqref{conjecture_Noether_1,5} could also holds for higher
rank BV theories.
\vspace{0.2cm}\\
- Appendix \ref{Annexe_constraints} details the constraints on the
parametrizing functions of the BRST transformations due to the nilpotency
of the BRST operator.
\vspace{0.2cm}\\
- Appendix \ref{Annexe_A} briefly recalls the BRST covariant phase space 
introduced in \cite{Baulieu:2024oql}, which is used throughout the article. 
\end{adjustwidth}

\section{BRST Noether current}\label{Section_BRST_Noether_current}

\subsection{The basic set-up}\label{section_set-up}

We consider a classical field theory on a $D$-dimensional spacetime $M$ with classical field space
$\mathcal{F}=\{\phi^i\}$ and a gauge invariant Lagrangian $L[\phi^i]$.
The BRST gauge-fixing procedure, necessary to 
define the perturbative quantum field theory associated to $L[\phi^i]$,  involves extending the classical fields $\phi^i$ to a BRST field multiplet 
\begin{equation} 
\label{BRST_multiplet}
\phi^i \to \Ph^I=(\phi^i,c^A,\bar{c}_A,b_A) ,
\end{equation}
where $c^A$ are the ghost fields associated with the gauge invariance,   $\bar{c}_A$ the antighost fields and $b_A$ the auxiliary fields.  The doublet $(\bar{c}_A,b_A)$ is cohomologically trivial. 
Importantly, the notion of gauge symmetry is defined on the classical fields $\phi^i$, not on the extended multiplet $\Ph^I$.  At the quantum level, it is replaced by BRST symmetry,  which generalizes gauge symmetry in a consistent manner.

Here we limit ourselves to rank-$1$ BV systems, namely to theories with an
off-shell closed gauge algebra. 
Such theories possess a graded off-shell nilpotent BRST operator~$s$ acting on \eqref{BRST_multiplet}, with a BV Lagrangian at most linear in the antifields. 
Accordingly,  the source terms in the generating functional are linearly coupled to the classical BRST variations of the fields, and their correlation functions are computed on the same footing as those of the fields themselves. This framework provides the natural setting in which the BRST Noether currents acquire a precise cohomological meaning and can be analyzed consistently at the quantum level.

The advantage of working with such rank-$1$ theories is that gauge fixing does not require antifields; instead,  one can simply add an $s$-exact term to the classical gauge-invariant Lagrangian. 
The gauge fixed and BRST invariant Lagrangian thus takes the form
\begin{equation}\label{L_GF}
\tilde{L}[\Ph^I]=L[\phi^i]+s(\Ps[\Ph^I]).
\end{equation}
It must have ghost number zero. 
The ghost number $-1$ local
functional~$\Ps$ characterizes the chosen gauge fixing for the fields.

It must be appreciated that the BRST fields $\Ph^I$ and the residual BRST symmetry of~\eqref{L_GF} are essential
to defining the unitary $\mathcal{S}$-matrix of any given gauge theory.  
Physical $\ket{\rm in}$ and $\ket{\rm out}$ states are also unambiguously
defined from the cohomology of the nilpotent BRST operator $s$ associated
with the gauge symmetry. 

We now describe in detail the class of theories we consider.
In the context of asymptotic symmetries, we wish to establish general formulas
that are model independent.
We will thus use a rather general ansatz for the BRST transformations of
the various fields.
While it does not encompass all possibilities, it covers a sufficiently large class of rank-$1$ theories. 
Indeed,  it covers at least the cases of Yang-Mills theory
\cite{Strominger_2014,He:2015zea,Freidel:2023gue,Nagy:2024jua},
general relativity in its first and second order formulations 
\cite{bondi,sachs,BMS/CFT,Campiglia,Compere,Oliveri:2019gvm,Freidel1,Geiller,
Baulieu_Tom_BMS}, theories involving coupled abelian $2$-forms
\cite{Afshar:2018apx,Campiglia:2018see,Francia:2018jtb,Romoli:2024hlc},
$\mathcal{N}=1$, $d=4$ supergravity with auxiliary fields
\cite{Awada:1985by,Avery:2015iix,Fotopoulos:2020bqj,Fuentealba:2021xhn,
Tomova:2022caw,Bandos:2024pns} and the Poisson BF theory in twistor space
\cite{Adamo:2021lrv,Donnay:2024qwq,Kmec:2024nmu} that leads to self-dual
gravity \cite{Strominger:2021mtt,Ball:2021tmb,Cresto:2024mne}. 
The classical canonical generators of the asymptotic symmetries of such
theories are Noether charges.
Extending Noether's second theorem for the gauge fixed version of these
theories is therefore essential. 
We also include super Yang-Mills theory
\cite{Dumitrescu:2015fej,Pano:2021ewd}, the worldsheet theory of strings and
superstrings, as well as all possible couplings between these theories and
the previous ones. 
 
Since ghosts of ghosts phenomena often occur, for instance in supergravity,
we need to distinguish ghost number one fields $c^A_1$ with ghost number two 
fields $c^P_2$.  
Indices $\{A,B,C,\dots\}$ will consistently label $c_1$, while indices
$\{P,Q,R,\dots\}$ label $c_2$.
The field space is thus extended to the space of
$\Ph^I=(\phi^i,c^A_1,c^P_2,\bar{c}^{\ -1}_A,b^0_A,\bar c^{\ -2}_P,b^{-1}_P)$,
where the numerical subscripts or superscripts indicate the ghost numbers of
the fields.
For simplicity, we adopt the following slight abuse of notation:
$\bar{c}_I=(\bar{c}^{\ -1}_A,\bar{c}^{\ -2}_P)$ and $b_I=(b^0_A,b^{-1}_P)$.
To understand how all these indices are used in practical computations, see
Section~\ref{Section_examples}.  
Since we won't consider $p$-form gauge theories with $p>2$,  we do not need
ghosts with a ghost number higher than $2$.

We propose the following ansatz for the rather general nilpotent BRST
transformations covering at least all rank-$1$ BV cases mentioned above: 
\begin{align}
\label{BRST_trans}
s\phi^i&=f^i_A(\phi,\pa_\nu\phi)c^A_1+f^{\mu i}_A(\phi)\pa_\mu c^A_1,    \nn\\
sc^A_1&=\gamma^A_{\ P}(\phi)c^P_2+M^{\mu A}_P(\phi)\pa_\mu c^P_2
+\gamma^A_{\ BC}(\phi)c^B_1c^C_1+M^{\mu A}_{BC}(\phi)c^B_1\pa_\mu c^C_1, \nn\\
sc^P_2&=\beta^P_{\ ABC}(\phi)c^A_1c^B_1c^C_1+\varGamma^P_{\ QA}(\phi)
 c^Q_2c^A_1+\varGamma^{\mu P}_{AQ}(\phi)c^A_1\pa_\mu c^Q_2,              \nn\\
s\bar{c}_I&=b_I,                                                         \nn\\
sb_I&=0.
\end{align}
The BRST transformation $s\phi^i$ comes from the gauge transformations of the
classical Lagrangian $L[\phi^i,\pa_\nu\phi^i]$.
All unspecified functions of the fields in this parametrization are assumed to
be polynomials.
We also assume that $f^i_A(\phi,\pa_\nu\phi)$ is the sum of a polynomial in
$\phi$ and a linear term in $\pa_\nu\phi$ with a constant coefficient. 
For the sake of notational simplicity, this paper only gives detailed formula
for cases where the classical fields $\phi^i$ are commuting ones so that the
ghost fields $c^A_1$ are anticommuting and the ghost fields $c^P_2$ are
commuting.
This is not a lack of generality and our formula can be easily generalized,
for example, to supergravity and other theories involving fermionic matter.  

The consequences of the nilpotency of $s$ on our parametrizing field dependent
functions $f^i_A,f^{\mu i}_A,\gamma^A_{\ P},M^{\mu A}_P,\gamma^A_{\ BC},
M^{\mu A}_{BC},\beta^P_{\ ABC},\varGamma^P_{\ QA}$ and
$\varGamma^{\mu P}_{AQ}$
are essential.\footnote{For the most general Taylor expansion of the BRST-BV
operator $s$, containing both fields and antifields, it is known that the
nilpotency condition $s^2=0$ defines an $L_\infty$ structure, see for example
\cite{Alexandrov:1995kv, Kontsevich:1997vb,Grigoriev:2023lcc}.
We will certainly need this structure to extend the proof of the BRST Noether
theorem to all rank BV systems.
Here we choose to restrict ourselves to the expansion \eqref{BRST_trans} in
order to obtain explicit expressions for a large enough class of theories.} 
To work out these results, we expand $s^2\Ph^I$ in a polynomial basis for the
ghosts and their derivatives and impose that each component vanish.
On the classical fields $\phi^i$, ghosts $c^A_1$, and ghosts of ghosts
$c_2^P$, we find 
\begin{align}\label{s2cphi}
0=s^2\phi^i
&=c^P_2\big(X^i_1\big)_P+\pa_\mu\pa_\nu c^P_2\big(X^i_2\big)^{\mu\nu}_P
 +\pa_\mu c^P_2\big(X^i_3\big)_P^\mu+c^B_1c^A_1\big(X^i_4\big)_{BA}
 +\pu c^B_1c^A_1\big(X^i_5\big)_{BA}^\mu \nn \\
&\quad+\pu\pv c^B_1c^A_1\big(X^i_6\big)_{BA}^{\mu\nu}
 +\pv c^B_1\pu c^A_1\big(X^i_7\big)_{BA}^{\mu\nu} ,
\end{align}
\begin{align}\label{s2c1}
0=s^2c^A_1&=c^P_2c^D_1\big(Y^A_1\big)_{PD}
 +c^P_2\pa_\mu c^D_1\big(Y^A_2\big)_{PD}^\mu+c^D_1\pa_\mu\pa_\nu c^P_2
 \big(Y^A_3\big)_{PD}^{\mu\nu}+c^D_1\pu c^P_2\big(Y^A_4 \big)_{PD}^\mu  \nn\\
&\quad+\pa_\mu c^P_2\pa_\nu c^D_1\big(Y^A_5\big)_{PD}^{\mu\nu}
 +c^D_1c^B_1c^C_1\big(Y^A_6\big)_{DBC}
 +\pa_\mu c^D_1c^B_1c^C_1\big(Y^A_7\big)_{DBC}^\mu                      \nn\\
&\quad+\pu\pv c^D_1c^B_1c^C_1\big(Y^A_8\big)_{DBC}^{\mu\nu}
 +c^D_1\pv c^B_1\pu c^C_1\big(Y^A_9\big)_{DBC}^{\mu\nu},
\end{align}
\begin{align}\label{s2c2}
0=s^2c^P_2&=c^Q_2c^R_2\big(Z^P_1\big)_{QR}
 +c^Q_2\pa_\nu c^R_2\big(Z^P_2\big)_{QR}^\nu
 +\pa_\mu c^Q_2\pa_\nu c^R_2\big(Z^P_3\big)_{QR}^{\mu\nu}
 +c^Q_2c^C_1c^E_1\big(Z^P_4\big)_{QCE }                       \nn \\
&\quad+c^C_1c^D_1\pa_\mu\pa_\nu c^Q_2\big(Z^P_5\big)_{CDQ}^{\mu\nu}
 +c^B_1\pa_\mu c^Q_2\pa_\nu c^E_1\big(Z^P_6\big)_{BQE}^{\mu\nu}
 +c_2^Qc^C_1\pa_\nu c^E_1\big(Z^P_7\big)_{QCE}^\nu            \nn \\
&\quad+\pa_\mu c^Q_2c^B_1c^E_1\big(Z^P_8\big)_{QBE}^\mu
 +c^B_1c^C_1c^D_1c^E_1\big(Z^P_9\big)_{BCDE}
 +c^B_1c^C_1c^D_1\pa_\nu c^E_1\big(Z^P_{10}\big)_{BCDE}^\nu.
\end{align}
We computed all the $X^i$'s, $Y^A$'s and $Z^P$'s in \eqref{s2cphi},
\eqref{s2c1} and \eqref{s2c2} in terms of the parametrizing functions
defining the BRST transformations \eqref{BRST_trans}.
The vanishing of $X^i$'s, $Y^A$'s and $Z^P$'s leads to non-trivial
constraints that are exhibited in Appendix~\ref{Annexe_constraints}.  

We now turn to the determination of the Noether current associated with the BRST symmetry \eqref{BRST_trans} of the gauge fixed  Lagrangian $\tilde{L}$,  defined in \eqref{L_GF} in the $\Ps$ gauge.
For this purpose it is convenient to work in the trigraded BRST covariant phase space \cite{Baulieu:2024oql},  whose basic graded differentials are $d$, $\delta$ and  $s$.
These differentials are compatible, that is
\begin{equation}
\label{compatible_differential_CPS}
(d+\delta+s)^2=0.
\end{equation} 
This allows one to treat arbitrary infinitesimal variations of the fields $\delta \Phi^I$ in a unified way, as $1$-forms on field space. 
The total grading of each object is inherited from the  spacetime exterior derivative $d$,  the field space  exterior derivative  $\delta$ and the BRST operator $s$,  as well as the intrinsic gradings  of the  fields $\Ph^I$.  
Ghost number,  field space form degree and spacetime form degree are conserved throughout all equations,  as a guiding principle.

It must be noted that considerable mathematical depth has already been achieved in this trigraded covariant phase space framework.  Many results are well established, including the full BV extension formulated within the jet-bundle approach \cite{Grigoriev:2012xg}, which allows for arbitrary and well-defined generalizations of \eqref{BRST_trans} and incorporates the descent completion of the Lagrangian (the $K$ operator in \eqref{fundam_identities} and its higher analogs). The term ``tricomplex" has even already been used, e.g. in \cite{Sharapov:2016sgx}.

In the BRST covariant phase space,   the following fundamental identities   introduce the relevant local field operators~$\theta$,   $ \tilde \theta$,  $ \Xi$ and $K$,   which play a central role in the computation of the Noether current for the BRST symmetry:
\begin{align}\label{fundam_identities}
\delta L&=E_{\phi^i}\delta\phi^i+\pa_\mu\theta^\mu, & sL&=\pa_\mu K^\mu,\nn \\
\delta\tilde{L}&=\tilde{E}_{\Ph^I}\delta\Ph^I+\pa_\mu\tilde{\theta}^\mu,
& s\tilde{L}&=\pa_\mu K^\mu,      \nn \\
\delta\Ps&=F_{\Ph^I}\delta\Ph^I+\pa_\mu\Xi^\mu. 
\end{align}
The relation between $L$ and $\tilde{L}$ is explained in \eqref{L_GF}.  In the present context we work with spacetime multi-vector fields rather than spacetime differential forms, in order to facilitate various integrations by parts. Consequently, all quantities in \eqref{fundam_identities} are spacetime zero-forms.
They are nevertheless distinguished by their field space form degree and ghost number. 
Specifically,  the identities in the left column of \eqref{fundam_identities} have field space form degree one and ghost number zero,  whereas those in the right column have field space form degree zero and ghost number one.

The classical equations of motion $E_{\phi^i}$ and local symplectic potential $\theta^\mu$ are gauge independent. 
By contrast,  the equations of motion $\tilde{E}_{\Ph^I}$ and the total local symplectic
potential $\tilde{\theta}^\mu$ are gauge dependent; these are the relevant quantities for the gauge fixed theory.
In what follows, $\hat{=}$ denotes any equality that is valid only on-shell,
i.e., when the equations of motion $\tilde{E}_{\Ph^I}=0$ are satisfied.

All equations in \eqref{fundam_identities} remain valid and are sometimes more convenient to use
if we regard $L$ as a spacetime top-form by including a volume form factor $d^Dx$.  The corresponding local symplectic potential $\theta$ is then a spacetime $(D{-}1)$-form and field space $1$-form, defined by the contraction $\theta = \theta^\mu i_{\pa_\mu} (d^D x)$,  where $i_{\pa_\mu}$ is the spacetime interior product along the vector field $\pa_\mu$.  In this language the divergence $\pa_\mu \theta^\mu$ is simply $d\theta$.  The same applies to  $\pa_\mu \tilde{\theta}^\mu$, $\pa_\mu \Xi^\mu$ and $\pa_\mu K^\mu$. 
More generally,   a spacetime $k$-vector field $\zeta^{\mu_1 \cdots \mu_k}$ can be identified with the spacetime $(D{-}k)$-form 
\begin{equation}
\label{gamma_diff_form}
\zeta = \zeta^{\mu_1 \cdots \mu_k} i_{\pa_{\mu_1} \w \cdots \w \pa_{\mu_k} } (d^D x) .
\end{equation}
We will use this identification from time to time,  and when we do so it will be clear from context.

The identities \eqref{fundam_identities} imply constraints on the BRST transformation of the equations of motion, namely\footnote{
The function $\epsilon$ is the parity that determines the sign from commuting
quantities graded by the ghost number and the degree of forms on the field
space.}
\begin{align}\label{computation_Z}
\pa_\mu\delta K^\mu=&-s\delta\tilde{L} \nn \\
=&\ - s\tilde{E}_{\Ph^I}\delta\Ph^I+(-1)^{\epsilon(\tilde{E}_{\Ph^I})}
 \tilde{E}_{\Ph^I}\delta s\Ph^I-\pa_\mu(s\tilde{\theta}^\mu)  \nn \\
=&\ G_{\phi^j}\delta\phi^j+G_{c^B_1}\delta c^B_1+G_{c^P_2}\delta c^P_2
 +  G_{\bar{c}_I}\delta\bar{c}_I+G_{b_I}\delta b_I  \nn \\
&+\pa_\mu\Bigg(-s\tilde{\theta}^\mu+(-1)^{\epsilon(\tilde{E}_{\phi^i})}
 \tilde{E}_{\phi^i}\left[\frac{\pa f^i_B}{\pa\pa_\mu\phi^j}\delta\phi^jc^B_1
 +f^{\mu i}_B\delta c^B_1\right]  \nn \\
&+(-1)^{\epsilon(\tilde{E}_{c^A_1})}\tilde{E}_{c^A_1}
 \left[M^{\mu A}_P\delta c^P_2-M^{\mu A}_{BC}c^B_1\delta c^C_1\right]
 -(-1)^{\epsilon(\tilde{E}_{c^P_2})}
 \tilde{E}_{c^P_2}\varGamma^{\mu P}_{BQ}c^B_1\delta c^Q_2\Bigg).
\end{align}
This leads to the important functional constraints 
\begin{equation}
G_{\Ph^I}=0 .
\end{equation}
By working out the detailed expression of $G_{\Ph^I}$, one finds
$s\tilde{E}_{\Ph^I}\propto[\tilde{E}_{\Ph^J},c^C]^I$ so
$s\tilde{E}_{\Ph^I}\ \hat{=}\ 0$.
The expression of the boundary piece of \eqref{computation_Z} is to become
important when studying the non-integrable part of the gauge fixed
fundamental relation in Section~\ref{section_gauge_fixed_fund}. 

To determine precisely how $\tilde{E}_{\Ph^I}$ and $\tilde\theta^\mu$ depend
on the gauge fixing functional $\Ps$,  through its variation $\delta\Ps$,
we compute\footnote{The bracket $\{\cdots\}$ on indices $A,B, \dots$ denotes complete antisymmetrization with weight one.} 
\begin{align}
\label{theta_gauge}
\delta s \Ps &= - s \delta \Ps = - (s F_{\Ph^I} ) \delta \Ph^I + (-1)^{\epsilon(F_{\Ph^I})} F_{\Ph^I} \delta s \Ph^I - \pa_\mu (s \Xi^\mu)
\nn \\
&= \Bigg[  - s F_{\phi^i}  - (-1)^{\epsilon(F_{\phi^j})}  F_{\phi^j}  \frac{\pa}{\pa \phi^i} \left(  f^j_B  c^B_1  +   f^{\mu j}_B  \pa_\mu c^B_1 \right)  + (-1)^{\epsilon(F_{\phi^j})} \pa_\mu \left( F_{\phi^j} \frac{\pa f^j_B}{\pa \pa_\mu \phi^i} c^B_1 \right)
\nn \\
&\quad +  (-1)^{\epsilon(F_{c^A_1})}  F_{c^A_1}  \frac{\pa}{\pa \phi^i} \left(   \gamma^A_{\ P}  c^P_2 +  M^{\mu A}_P  \pa_\mu c^P_2 +  \gamma^A_{\ BC}  c^B_1 c^C_1 +   M^{\mu A}_{BC}  c^B_1 \pa_\mu c^C_1  \right)
\nn \\
&\quad - (-1)^{\epsilon(F_{c^P_2})}  F_{c^P_2}  \frac{\pa}{\pa \phi^i} \left(   \beta^P_{\ BCD}  c^B_1 c^C_1 c^D_1 +   \varGamma^P_{\ QC}  c^Q_2 c^C_1  +   \varGamma^{\mu P}_{BQ}  c^B_1 \pa_\mu c^Q_2  \right)  \Bigg] \delta \phi^i + \Bigg[   -s F_{c^A_1}
\nn \\
&\quad  + (-1)^{\epsilon(F_{\phi^i})} \left( F_{\phi^i} f^i_A - \pa_\mu (F_{\phi^i} f^{\mu i}_A) \right) + (-1)^{\epsilon(F_{c^B_1})} \Big( F_{c^B_1} ( \gamma^B_{\ \{AC\}}   c^C_1  + M^{\mu B}_{AC} \pa_\mu c^C_1  ) 
\nn \\
&\quad + \pa_\mu (F_{c^B_1} M^{\mu B}_{CA} c^C_1)  \Big)  + (-1)^{\epsilon(F_{c^P_2})}  F_{c^P_2} \left(  \beta^P_{\ \{ACD\}} c^C_1 c^D_1 + \varGamma^P_{\ QA} c^Q_2 + \varGamma^{\mu P}_{AQ} \pa_\mu c^Q_2  \right)   \Bigg] \delta c^A_1
\nn \\
&\quad + \Bigg[  -s F_{c^P_2} + (-1)^{\epsilon(F_{c^B_1})} \left( F_{c^B_1} \gamma^B_{\ P} - \pa_\mu ( F_{c^B_1} M^{\mu B}_P  ) \right) - (-1)^{\epsilon(F_{c^Q_2})}   \Big( F_{c^Q_2} \varGamma^Q_{\ PC} c^C_1
\nn \\
&\quad  - \pa_\mu ( F_{c^Q_2} \varGamma^{\mu Q}_{CP} c^C_1 ) \Big) \Bigg] \delta c^P_2 - \Big[ s F_{\bar{c}_I} \Big] \delta \bar{c}_I + \left[ - s F_{b_I} + (-1)^{\epsilon(F_{\bar{c}_I})} F_{\bar{c}_I} \right] \delta b_I
\nn \\
&\quad + \pa_\mu \Bigg[   - s \Xi^\mu +  (-1)^{\epsilon(F_{\phi^i})}   F_{\phi^i} \left(  \frac{\pa f^i_B}{\pa \pa_\mu \phi^j} \delta \phi^j c^B_1 + f^{\mu i }_B \delta c^B_1 \right) - (-1)^{\epsilon(F_{c^A_1})} F_{c^A_1} \Big(  M^{\mu A}_{BC} c^B_1 \delta c^C_1 
\nn \\
&\quad - M^{\mu A}_P \delta c^P_2  \Big) - (-1)^{\epsilon(F_{c^P_2})}  F_{c^P_2} \left(  \varGamma^{\mu P}_{BQ} c^B_1 \delta c^Q_2 \right)  \Bigg] .
\end{align}
The boundary term in this equation is denoted by $\theta^\mu_{\rm gauge}$.  It constitutes
the gauge dependent part of the local symplectic potential
$\tilde{\theta}^\mu$. 
The factors in front of $\delta c^A_1$, $\delta c^P_2$, $\delta \bar{c}_I$
and $\delta b_I$ are  the gauge dependent  equations of motion  
$\tilde{E}_{c^A_1}$, $\tilde{E}_{c^P_2}$, $\tilde{E}_{\bar{c}_I}$ and 
$\tilde{E}_{b_I}$, respectively.
The factor in front of $\delta\phi^i$ is the gauge dependent part
$\tilde{E}_{\phi^i}^{\rm gauge}$ of the equations of motion of the classical
fields $\phi^i$ defined in \eqref{fundam_identities}, and we have 
$\tilde{E}_{\phi^i}=E_{\phi^i}+\tilde{E}_{\phi^i}^{\rm gauge}$.

The missing ingredient for computing the complete BRST Noether current of the gauge fixed Lagrangian \eqref{L_GF} is the classical Noether current associated with the
gauge invariance of $L[\phi^i]$.
We understand from \eqref{fundam_identities} that this current satisfies
Noether's second theorem \cite{Noether:1918zz} and is given off-shell
by\footnote{More recent discussions of this theorem include
\cite{Avery_Schwab,Miller:2021hty}.} 
\begin{equation}\label{classical_J}
J^\mu_{\rm cl}=-E_{\phi^i}f^{\mu i}_Ac^A_1
 +\pa_\nu q^{\mu\nu}_{\text{cl},c^A_1}.
\end{equation}
The corner Noether charges $q^{\mu\nu}_{\text{cl},c^A_1}$ are linear in
$c^A_1$ and are the main focus of this paper since they probe asymptotic
symmetries and since their charge algebra projectively represents the
asymptotic symmetry algebra. 
The upshot of the BRST Noether 1.5th theorem is that we can still find these
charges when starting from a gauge fixed Lagrangian \eqref{L_GF}, and that
the physical implications of their existence are indeed gauge independent.

\subsection{The 1.5th theorem}\label{section_1.5}

We now state our main result.

\begin{theorem}[BRST Noether 1.5]
\label{claim_BRST_1.5}
Let $s$ be the off-shell nilpotent BRST operator \eqref{BRST_trans},  defining a class of rank-1 BV theories.  Consider the corresponding gauge fixed Lagrangian  
\begin{equation}\label{L_GF_thm}
\tilde{L}[\Ph^I]=L[\phi^i]+s(\Ps[\Ph^I]), 
\end{equation}
which is invariant under $s$.
 
Then,  on-shell of the gauge fixed equations of motion derived from this Lagrangian, 
    the ghost number one conserved Noether current $J^{\mu}_{\rm BRST}$ associated with the BRST invariance of~\eqref{L_GF_thm} decomposes as
\begin{equation}
J^{\mu}_{\rm BRST}\ \hat{=}\ sG_\Ps^\mu+\pa_\nu q^{\mu\nu},
\end{equation} 
where $q^{\mu\nu}$ is antisymmetric in $(\mu,\nu)$. 

The ghost number one corner charges $q^{\mu\nu}$ contain two distinct contributions:
\begin{itemize}
    \item[(i)] terms proportional to the primary ghost fields $c_1^A$ with coefficients depending only on the classical fields~$\phi^i$, which define the classical Noether charges $q^{\mu\nu}_{\mathrm{cl},\,c_1^A}$,
    \item[(ii)] gauge dependent higher-order terms involving ghosts and antighosts, collectively denoted by $q^{\mu\nu}_\Psi$.
\end{itemize}

The $s$-exact term,  determined explicitly by a forthcoming general
computation of the gauge dependent ghost number zero local functional
$G_\Ps^\mu[\Ph^I]$,  is another gauge dependent extension of the classical
Noether second theorem.
\end{theorem}

\begin{proof}
We work in the set-up of Section~\ref{section_set-up}, so that all assumptions of the theorem are satisfied. 
For the purpose of this proof, it is convenient to introduce a geometric definition of the BRST Noether current, based solely on the objects defined in \eqref{fundam_identities}. To this end, we first introduce additional structures of the BRST covariant phase space \cite{Baulieu:2024oql}.

In the BRST covariant phase space, the action of $s$ on higher field space
forms is inherited  from its action on field space zero forms \eqref{BRST_trans},  namely on the
fields $\Ph^I$, by considering the ghost number one field space vector field
\begin{equation}\label{V_BRST_main}
V_{\text{\tiny BRST}}\equiv\int_M d^Dx \ s\Ph^I(x)\frac{\delta}{\delta\Ph^I(x)}. 
\end{equation}
From there, the nilpotent BRST operator acting on the whole covariant phase space and
satisfying the compatibility condition \eqref{compatible_differential_CPS} is defined as
\begin{equation}\label{defining_s_main}
s\equiv L_{V_{\text{\tiny BRST}}}=I_{V_{\text{\tiny BRST}}}\delta-\delta I_{V_{\text{\tiny BRST}}}.
\end{equation}
Here $L_{V_{\text{\tiny BRST}}}$ and $I_{V_{\text{\tiny BRST}}}$ are the field space Lie derivative and field space interior product along the specific vector field
\eqref{V_BRST_main},  respectively.  The unexpected
minus sign in the definition of the Lie derivative comes from the odd grading
of $V_{\text{\tiny BRST}}$ for its ghost number.\footnote{In the language of \cite{Mnev:2019ejh},  $V_{\text{\tiny BRST}}$  is the degree one evolutionary and cohomological field space vector field defining a lax BV--BFV theory. }
When acting on the fields, the definition \eqref{defining_s_main} naturally reduces to \eqref{BRST_trans} since 
\begin{equation}
L_{V_{\text{\tiny BRST}}} \Ph^I = I_{V_{\text{\tiny BRST}}}\delta \Ph^I = s \Ph^I,
\end{equation}
as a consequence of \eqref{V_BRST_main}.

Having established these identities, we can now apply $I_{V_{\text{\tiny BRST}}}$ to $\delta \tilde{L}$ in \eqref{fundam_identities}. Combining this with $\tilde{E}_{\phi^i}=E_{\phi^i}+\tilde{E}_{\phi^i}^{\rm gauge}$ and \eqref{classical_J},
the conserved BRST Noether current associated with the Lagrangian~\eqref{L_GF_thm} takes the form
\begin{equation}
J^\mu_{\rm BRST}=I_{V_{\text{\tiny BRST}}}\tilde{\theta}^\mu - K^\mu=-E_{\phi^i}f^{\mu i}_Ac^A_1
 +\pa_\nu q^{\mu\nu}_{\text{cl},c^A_1}+I_{V_{\text{\tiny BRST}}} \theta^\mu_{\rm gauge}.
\end{equation}
A detailed derivation showing that this expression indeed defines the conserved BRST Noether current is provided in Appendix~\ref{Annexe_A}. Using the explicit expression of $\theta^\mu_{\mathrm{gauge}}$ given in~\eqref{theta_gauge}, this current can be further rewritten as
\begin{align}\label{full_BRS_current}
J^{\mu}_{\rm BRST} &= - \tilde{E}_{\phi^i} f^{\mu i}_A c^A_1 + \pa_\nu q^{\mu\nu}_{\text{cl},c^A_1} - s \left(   I_{V_{\text{\tiny BRST}}} \Xi^\mu + F_{\phi^i} f^{\mu i}_A c^A_1 \right) 
\nn \\
&\quad
+ (-1)^{\epsilon(F_{\phi^i})}   F_{\phi^i} \Bigg(  \left(\frac{\pa f^{\mu i}_A}{\pa \phi^j} + \frac{\pa f^i_A}{\pa \pa_\mu \phi^j} \right)s \phi^j c^A_1  + 2 f^{\mu i}_A sc^A_1 \Bigg)
\nn \\
&\quad  - (-1)^{\epsilon(F_{\phi^j})} \Bigg(  F_{\phi^j} \left( \frac{\pa f^j_B}{\pa \phi^i} c^B_1 + \frac{\pa f^{\nu j}_B}{\pa \phi^i} \pa_\nu c^B_1 \right) - \pa_\nu \left(  F_{\phi^j}  \frac{\pa f^j_B}{\pa \pa_\nu \phi^i} c^B_1 \right)   \Bigg) f^{\mu i}_A c^A_1
\nn \\
&\quad + (-1)^{\epsilon(F_{c^B_1})}  F_{c^B_1} \left(  \frac{\pa \gamma^B_{\ P}}{\pa \phi^i} c^P_2 + \frac{\pa M^{\nu B}_P}{\pa \phi^i} \pa_\nu c^P_2 + \frac{\pa \gamma^B_{\ CD}}{\pa \phi^i} c^C_1 c^D_1 + \frac{\pa M^{\nu B}_{CD}}{\pa \phi^i} c^C_1 \pa_\nu c^D_1 \right) f^{\mu i}_A c^A_1
\nn \\
&\quad - (-1)^{\epsilon(F_{c^P_2})}  F_{c^P_2} \left( \frac{\pa \beta^P_{\ CDE}}{\pa \phi^i} c^C_1 c^D_1 c^E_1 + \frac{\pa \varGamma^P_{\ QD}}{\pa \phi^i} c^Q_2 c^D_1 + \frac{\pa \varGamma^{\nu P}_{CQ}}{\pa \phi^i} c^C_1 \pa_\nu c^Q_2 \right) f^{\mu i}_A c^A_1
\nn \\
&\quad - (-1)^{\epsilon(F_{c^B_1})}  F_{c^B_1} M^{\mu B}_{CA} c^C_1 sc^A_1 + \left( (-1)^{\epsilon(F_{c^B_1})}  F_{c^B_1} M^{\mu B}_{P} -  (-1)^{\epsilon(F_{c^Q_2})}  F_{c^Q_2} \varGamma^{\mu Q}_{CP} c^C_1 \right) sc^P_2 .
\end{align}

Since the three terms in the first line of \eqref{full_BRS_current} leads to
the desired form of Theorem~\ref{claim_BRST_1.5}, we only have to focus on the other
ones and analyze their $F_{\Ph^I}$ dependence.
Let us start by computing the $F_{\phi^i}$ dependent part
of~\eqref{full_BRS_current}.
Equations \eqref{27}, \eqref{28}, \eqref{210}, \eqref{211}, \eqref{212} and
further integration by parts allow us to express this part of the Noether
current as
\begin{multline}\label{J_F_cl}
J^\mu_{\rm BRST}\supset(-1)^{\epsilon(F_{\phi^i})}\pa_\nu
\Bigg(F_{\phi^i}\left[c^B_1c^A_1\Big(\frac{\pa f^i_B}{\pa\pa_\nu\phi^j}
 f^{\mu j}_A-M^{\mu C}_{BA}f^{\nu i}_C\Big)-2c^P_2M^{\mu B}_Pf^{\nu i}_B
 \right]\Bigg)                                                          \\
\quad+(-1)^{\epsilon(F_{\phi^i})}\Big(F_{\phi^i}f^i_C
 -\pa_\nu(F_{\phi^i}f^{\nu i}_C)\Big)\Big(c^B_1c^A_1M^{\mu C}_{AB}
 -2c^P_2M^{\mu C}_P\Big).
\end{multline}
The boundary term is antisymmetric in $(\mu,\nu)$ by \eqref{27} and
\eqref{211} and is part of $q^{\mu\nu}$.
The non-boundary term is proportional to the $F_{\phi^i}$ dependent piece of
the equations of motion of $c^C_1$ in \eqref{theta_gauge}.  
Therefore, it can only produce terms that are on-shell proportional to
$F_{c^A_1}$ and~$F_{c^P_2}$.  

We must then  proceed to the evaluation of the $F_{c^A_1}$ dependent terms of
the BRST Noether current.
By \eqref{theta_gauge}, \eqref{full_BRS_current}, \eqref{J_F_cl} and one
integration by parts, we obtain that these terms are 
\begin{align}\label{J_Fc1}
J^{\mu}_{\rm BRST} &\supset s \Big( F_{c^C_1} [ c^B_1 c^A_1 M^{\mu C}_{AB} - 2 c^P_2 M^{\mu C}_P  ] \Big) - (-1)^{\epsilon(F_{c^D_1})} \pa_\nu \Big( F_{c^D_1} M^{\nu D}_{EC} c^E_1 [ c^B_1 c^A_1 M^{\mu C}_{AB} - 2 c^P_2 M^{\mu C}_P  ] \Big)
\nn \\
&\quad +  (-1)^{\epsilon(F_{c^C_1})}  F_{c^C_1} \bigg(  -2 M^{\mu C}_{AB}  c^A_1 sc^B_1 + M^{\mu C}_{AB} sc^A_1 c^B_1   -  \frac{\pa M^{\mu C}_{AB}}{\pa \phi^i} s \phi^i c^B_1 c^A_1   +3  M^{\mu C}_P sc^P_2  
\nn \\
&\quad +2 \frac{\pa M^{\mu C}_P}{\pa \phi^i} s \phi^i c^P_2   \bigg) +  (-1)^{\epsilon(F_{c^D_1})}  F_{c^D_1} M^{\nu D}_{EC} c^E_1 \pa_\nu \Big( c^B_1 c^A_1 M^{\mu C}_{AB} - 2 c^P_2 M^{\mu C}_P  \Big)
\nn \\
&\quad -   (-1)^{\epsilon(F_{c^D_1})}  F_{c^D_1} \Big(  \gamma^D_{\ \{CE\}} c^E_1 + M^{\nu D}_{CE} \pa_\nu c^E_1  \Big) \Big( c^B_1 c^A_1 M^{\mu C}_{AB} - 2 c^P_2 M^{\mu C}_P  \Big)
\nn \\
&\quad + (-1)^{\epsilon(F_{c^B_1})}  F_{c^B_1} \left(  \frac{\pa \gamma^B_{\ P}}{\pa \phi^i} c^P_2 + \frac{\pa M^{\nu B}_P}{\pa \phi^i} \pa_\nu c^P_2 + \frac{\pa \gamma^B_{\ CD}}{\pa \phi^i} c^C_1 c^D_1 + \frac{\pa M^{\nu B}_{CD}}{\pa \phi^i} c^C_1 \pa_\nu c^D_1 \right) f^{\mu i}_A c^A_1 . 
\end{align}
The first two terms in the first line  have the correct form according to Theorem~\ref{claim_BRST_1.5}. 
We thus ignore them from now on and focus on the rest of the terms.
The later can be grouped by their monomial dependence on the ghost fields and
their derivatives as in \eqref{s2c1}. 
We obtain the following cancellations.
The term proportional to $c^A_1 c^D_1 c^E_1$ vanishes by \eqref{219}.
The term proportional to $c^A_1c^D_1\pa_\nu c^E_1$ vanishes by \eqref{220}
and \eqref{221}.
By using \eqref{215} and \eqref{217} in the terms proportional to
$c^A_1 \pa_\nu c^P_2$, \eqref{214} and \eqref{216} in the terms proportional
to $c^P_2  c^B_1$, and \eqref{217} and one integration by parts in the terms
proportional to $c^P_2 \pa_\nu c^B_1$, we can treat all the terms in
\eqref{J_Fc1} and we are left with
\begin{multline}\label{J_Fc_leftover}
J^\mu_{\rm BRST}\supset-(-1)^{\epsilon(F_{c^C_1})}\pa_\nu
\Big(2F_{c^C_1}M^{\nu C}_Q\varGamma^{\mu Q}_{BP}c^B_1c^P_2\Big)      \\
\quad+2(-1)^{\epsilon(F_{c^C_1})}\Big(\pa_\nu(F_{c^C_1}M^{\nu C}_Q)
 -F_{c^C_1}\gamma^C_{\ Q}\Big)\varGamma^{\mu Q}_{BP}c^B_1c^P_2. 
\end{multline}
When combining the boundary term of \eqref{J_Fc1} with that of
\eqref{J_Fc_leftover}, we get a new contribution to $q^{\mu\nu}$ which is
indeed antisymmetric in $(\mu,\nu)$ by \eqref{215} and \eqref{220}.  
The non-boundary term of \eqref{J_Fc_leftover} is proportional to the
$F_{c^C_1}$ dependent part of the equations of motion of 
$c^Q_2$ in~\eqref{theta_gauge}.
This means that when one goes on-shell, all terms that remain are
proportional to $F_{c^P_2}$. 

Thus the final task consists of analyzing the terms proportional to
$F_{c^P_2}$ in the BRST Noether current \eqref{full_BRS_current}.
To do so, we extract the corresponding terms in \eqref{full_BRS_current},
we go on-shell in the non-boundary terms of \eqref{J_F_cl} and
\eqref{J_Fc_leftover} and we perform one more integration by parts.
We then obtain  
\begin{align}\label{J_Fc2}
J^\mu_{\rm BRST}
&\supset-s\Big(2 F_{c^P_2}c^D_1c^Q_2\varGamma^{\mu P}_{DQ}\Big)
 +(-1)^{\epsilon(F_{c^P_2})}\pa_\nu\Big(2F_{c^P_2}c^C_1c^D_1c^Q_2 
  \varGamma^{\nu P}_{CR}\varGamma^{\mu R}_{DQ}\Big)                    \nn \\
&\quad+(-1)^{\epsilon(F_{c^P_2})}F_{c^P_2}\bigg(2sc^D_1c^Q_2
  \varGamma^{\mu P}_{DQ}-3c^B_1sc^Q_2\varGamma^{\mu P}_{BQ}
 -2c^D_1c^Q_2\frac{\pa\varGamma^{\mu P}_{DQ}}{\pa\phi^i}s\phi^i        \nn \\
&\quad-2c^C_1\pa_\nu(c^D_1c^Q_2\varGamma^{\mu R}_{DQ})\varGamma^{\nu P}_{CR}
 -2c^C_1c^D_1c^Q_2\varGamma^P_{\ RC}\varGamma^{\mu R}_{DQ}\bigg)       \nn \\
&\quad-(-1)^{\epsilon(F_{c^P_2})}F_{c^P_2}\Big(\beta^P_{\ \{ACD\}} c^C_1c^D_1
 +\varGamma^P_{\ QA}c^Q_2+\varGamma^{\nu P}_{QC}\pa_\nu c^Q_2\Big)
  \Big(c^F_1c^E_1M^{\mu A}_{EF}-2c^R_2M^{\mu A}_R\Big)                 \nn \\
&\quad-(-1)^{\epsilon(F_{c^P_2})}F_{c^P_2}
 \left(\frac{\pa\beta^P_{\ CDE}}{\pa \phi^i}c^C_1c^D_1c^E_1
  +\frac{\pa\varGamma^P_{\ QD}}{\pa\phi^i}c^Q_2c^D_1
  +\frac{\pa\varGamma^{\nu P}_{CQ}}{\pa\phi^i}c^C_1\pa_\nu c^Q_2\right)
   f^{\mu i}_Ac^A_1. 
\end{align} 
The first two terms have the right form.
In particular, the boundary term either vanishes or is antisymmetric in
$(\mu,\nu)$ by \eqref{226} and is therefore part of $q^{\mu\nu}$.
All the other terms, which are proportional to $c^C_1c^D_1c^F_1c^E_1$,
$c^C_1\pa_\nu c^D_1c^P_2$, $c^C_1c^D_1\pa_\nu c^P_2$, $c^C_1c^D_1c^P_2$,
$c^P_2c^Q_2$ and $\pa_\nu c^P_2c^Q_2$, identically vanish by \eqref{231},
\eqref{227}, \eqref{226}-\eqref{227}, \eqref{228}-\eqref{229}, \eqref{223}
and \eqref{224}, respectively.

Putting everything together, we get
\begin{align}\label{Noether_1.5}
J^\mu_{\rm BRST} =& \ sG_\Ps^\mu
 +\pa_\nu\left(q^{\mu\nu}_{\text{cl},c^A_1}+q^{\mu\nu}_\Ps\right) - \tilde{E}_{\phi^i} f^{\mu i}_A c^A_1
 \nn \\  
 &+ \tilde{E}_{c^C_1} \left( c^B_1 c^A_1 M^{\mu C}_{AB} - 2 c^P_2 M^{\mu C}_P  \right) - 2 \tilde{E}_{c^P_2} c^D_1 c^Q_2 \Gamma^{\mu P}_{DQ} 
\nn \\
\hat{=}& \ sG_\Ps^\mu
 +\pa_\nu\left(q^{\mu\nu}_{\text{cl},c^A_1}+q^{\mu\nu}_\Ps\right) ,
\end{align}  
with
\begin{align}\label{explict_gauge_dependence}
\bullet \ G_\Ps^\mu &= - I_{V_{\text{\tiny BRST}}} \Xi^\mu - F_{\phi^i} f^{\mu i}_A c^A_1 + F_{c^C_1} \left( c^B_1 c^A_1 M^{\mu C}_{AB} - 2 c^P_2 M^{\mu C}_P  \right) - 2  F_{c^P_2} c^D_1 c^Q_2  \varGamma^{\mu P}_{DQ} 
\nn \\
\bullet \ q^{\mu\nu}_\Ps &=  (-1)^{\epsilon(F_{\phi^i})}    F_{\phi^i} \left( c^B_1 c^A_1 \Big( \frac{\pa f^i_B}{\pa \pa_\nu \phi^j} f^{\mu j}_A - M^{\mu C}_{BA} f^{\nu i}_C \Big) - 2 c^P_2 M^{\mu B}_P f^{\nu i}_B  \right)     \nn \\ 
&\quad  - (-1)^{\epsilon(F_{c^C_1})}  F_{c^C_1} \bigg(  c^E_1  c^B_1 c^A_1 M^{\nu C}_{ED} M^{\mu D}_{AB} + 2 c^B_1 c^P_2 \Big( M^{\nu C}_Q \varGamma^{\mu Q}_{BP}  -M^{\nu C}_{BD} M^{\mu D}_P  \Big)   \bigg)  \nn \\
&\quad + 2 (-1)^{\epsilon(F_{c^P_2})}    F_{c^P_2} c^C_1 c^D_1 c^Q_2 \varGamma^{\nu P}_{CR} \varGamma^{\mu R}_{DQ}\ .
\end{align}
This completes the proof of Theorem~\ref{claim_BRST_1.5}. 
\end{proof}

Notice that the only terms that are linear in the ghost field $c^A_1$ are
either in the $s$-exact piece or in the classical Noether charge
$q^{\mu\nu}_{\text{cl},c^A_1}$.
This will become very important when analyzing the physical consequences of
\eqref{Noether_1.5} in Section~\ref{Holographic_WI}.
One can also check that \eqref{explict_gauge_dependence} correctly reproduces
the BRST Noether currents found in \cite{Baulieu:2024oql} for Yang-Mills
theory and gravity in various gauges.

Eq.~\eqref{Noether_1.5} can equivalently be expressed in the language of spacetime differential forms:
\begin{equation}
\label{Noether_1.5_form}
J_{\rm BRST} \ \hat{=} \  s G_\Ps + d ( q_{\text{cl},c^A_1}+q_\Ps ),
\end{equation}
where $J_{\rm BRST}$,  $G_\Ps$,  $q_{\text{cl},c^A_1}$ and $q_\Ps$ are defined by $J^\mu_{\rm BRST}$,  $G^\mu_\Ps$,  $q^{\mu \nu}_{\text{cl},c^A_1}$ and $q^{\mu \nu}_\Ps$
as in \eqref{gamma_diff_form}, respectively. 

Let us close this section with a simple example of a 
gauge fixing functional~$\Ps$.
If one chooses gauge fixing functionals $\mathcal{F}^A(\phi^i)$
that are independent of the field derivatives and of the ghosts, i.e.,
\begin{equation}\label{psi_LC}
\Ps=\bar{c}_A\mathcal{F}^A(\phi^i),
\end{equation}
such as in the Yang-Mills axial gauge, one gets
\begin{equation}
\delta\Ps=\delta\bar{c}_A\mathcal{F}^A
 -\bar{c}_A\frac{\pa\mathcal{F}^A}{\pa\phi^i}\delta\phi^i.
\end{equation}
In this case, one has $F_{c^A_1}=F_{c^P_2}=F_{b_A}=0$,
$F_{\bar{c}_A}=\mathcal{F}^A$ and
$F_{\phi^i}=-\bar{c}_A\frac{\pa\mathcal{F}^A}{\pa\phi^i}$.
The equations of motion of $b_A$ in \eqref{theta_gauge} then imply 
\begin{equation}\label{constraints_specific_gauge}
F_{\bar{c}_A}\ \hat{=}\ 0\quad\Longrightarrow\quad
\bar{c}_A\delta F_{\bar{c}_A}\ \hat{=}\ 0\quad\Longrightarrow\quad
F_{\phi^i}\phi^i\ \hat{=}\ 0. 
\end{equation}

All this means that for gauge fixing functionals of the type \eqref{psi_LC},
the explicit expressions in \eqref{explict_gauge_dependence} drastically
simplify as the terms in $F_{c^A_1}$ and $F_{c^P_2}$ drop out.
One can also make use of \eqref{constraints_specific_gauge} to further
simplify the $F_{\phi^i}$ dependent part of \eqref{explict_gauge_dependence}.
This last step can only be done after having specified the details of the
theory, namely its field content and the nilpotent BRST transformations
\eqref{BRST_trans}.

Such restricted solutions for the BRST Noether 1.5th theorem are relevant
since they appear for example when studying asymptotically flat gravity in
the partial Bondi gauge \cite{Geiller,Geiller:2024amx}, in which the
computations of \cite{Baulieu:2024oql} have been performed and which we know
leads to non-trivial asymptotic symmetries.
In this case, one finds $q^{\mu\nu}_{\Ps,{\rm Bondi}}\ \hat{=}\ 0$.
Rather, if we work in the de~Donder gauge, we find by using
\eqref{explict_gauge_dependence} and the equations of motion
$\tilde{E}_{b^\mu}$ that
$q^{\mu\nu}_{\Ps,\text{de\ Donder}}\ \hat{=}\ 
-\xi^\beta\xi^{\{\nu}\pa^{\mu\}}\big({\bar{\xi}}^\alpha g_{\alpha \beta})\ne0$
even on-shell.\footnote{The $\xi^\mu$'s are the vector primary ghosts for
reparametrization symmetry, the ${\bar{\xi}}^\mu$'s their antighosts, and
$b^\mu$ is the auxiliary field such that $s\bar\xi^\mu=b^\mu$ and
$sb^\mu=0$.}

Another example, that of an abelian $2$-form coupled to Chern-Simons theory in
four-dimensions, will be explained in Section~\ref{Section_examples} to help
the reader gain intuition about the various abstract notations of this
section.

\subsection{Holographic Ward identities}\label{Holographic_WI}

The BRST Noether currents found in \cite{Baulieu:2024oql} for specific
examples don't exhibit any $q^{\mu\nu}_\Ps$ dependence.
However, in theories such as gravity, such terms appear in the de~Donder
gauge.
Here we show that these gauge dependent charges do not spoil the gauge
independence of the Hamiltonian Ward identity $[Q_\lambda,\mathcal{S}]=0$,
thus providing a general proof that the physical $\mathcal{S}$-matrix must be
invariant under asymptotic symmetries \cite{Strominger_BMS_scattering}.
This Ward identity is non-trivial in theories with massless gauge particles
and is actually equivalent to the soft theorems for these particles
\cite{Strominger_lectures}. 

According to Noether second theorem,  asymptotic symmetries are present whenever there exist non-vanishing charges 
\begin{equation}\label{corner_charge}
Q_{\lambda^A}\equiv\int_{\pa\varSigma}q_{\text{cl},c^A_1}
\Big\vert_{c^A_1=\lambda^A}\ne0
\end{equation}
associated with a gauge symmetry parameter $\lambda^A(x)$ defined on the
boundary of a codimension-one Cauchy hypersurface $\varSigma$.
The ghost number one charges $q_{\text{cl},c^A_1}$ are defined in~\eqref{Noether_1.5_form}. 
They play the role of a substitute for the classical ghost number zero Noether charges $q_{\lambda^A}$, which are obtained by replacing
  the primary ghosts $c^A_1$ of the BRST symmetry with their associated infinitesimal gauge parameters $\lambda^A$, as in \eqref{corner_charge}.
The previous section has thus established \eqref{corner_charge} as a  generalization of the classical Noether second theorem using a vanishing gauge fixing function $\Ps$.
 
Now to demonstrate that Strominger's Hamiltonian postulate $[Q_{\lambda^A},\mathcal{S}]=0$ holds in a gauge fixed
theory independently of the value of $q_\Ps$, we rely on the boundary BRST
Ward identities derived in \cite{Baulieu:2024oql}.
These are all Ward identities for connected one-particle-irreducible
Green functions built from the gauge fixed and BRST invariant Lagrangian~\eqref{L_GF_thm},  with insertion of a BRST Noether current \eqref{Noether_1.5_form} on the null boundaries $\mathcal{I}^\pm$ of a flat or asymptotically flat spacetime $M$. 
They take the following form \cite{Baulieu:2024oql} 
\begin{multline}\label{Noether_ward_identity}
\left\langle\left(\int_{\mathcal{I}^+}J_{\text{BRST}}(x)\right)\mathcal{O}_1(\phi^{i_1}(x_1))
 \cdots\mathcal{O}_n(\phi^{i_n}(x_n))\right\rangle                            \\
-\left\langle\mathcal{O}_1(\phi^{i_1}(x_1))\cdots\mathcal{O}_n((\phi^{i_n}(x_n))
 \left(\int_{\mathcal{I}^-}J_{\text{BRST}}(x)\right)\right\rangle=0,
\end{multline}
which expresses the conservation of the BRST Noether current 
in any scattering process of massless particles propagating from $\mathcal{I}^-$ to $\mathcal{I}^+$.
The physical observables $\mathcal{O}_i$ are inserted at the bulk points $x_1, \dots, x_n$ and belong to the cohomology of the BRST
operator at ghost number zero,  hence they depend only on the classical fields $\phi^i$.  

Let us now analyze the
implications of \eqref{Noether_ward_identity} for asymptotic symmetries. Their action at future null infinity is generated by the charges \eqref{corner_charge} when the Cauchy hypersurface is chosen to be $\mathcal{I}^+$~(respectively $\mathcal{I}^-$ for past null infinity). 
On-shell, these charges satisfy flux-balance laws: they decrease (increase) along $\mathcal{I}^+$ ($\mathcal{I}^-$). We can therefore impose that they vanish at the future (past) boundary of $\mathcal{I}^+$ ($\mathcal{I}^-$),  namely at 
$\mathcal{I}^+_+$ ($\mathcal{I}^-_-$).  Consequently, only the boundaries $\mathcal{I}^+_-$ and $\mathcal{I}^-_+$ of future and past null infinities contribute in what follows.

We shall now explain why the possible dependence of the BRST Noether current
$\J$ on the gauge fixing functional $\Ps$ is irrelevant for physical
consequences of asymptotic symmetries, as it must be.
To do so, we combine \eqref{Noether_ward_identity} with the 1.5th theorem
\eqref{Noether_1.5_form} that expresses the BRST Noether current as
$J_{\rm BRST} \ \hat{=} \  s G_\Ps + d ( q_{\text{cl},c^A_1}+q_\Ps )$. 
It provides
\begin{equation}\label{WI_q_gauge}
\left\langle\int_{\mathcal{I}^+_-}\Big(q_{\text{cl},c^A_1}+q_\Ps\Big)
\mathcal{O}_1\cdots\mathcal{O}_n\right\rangle
-\left\langle\mathcal{O}_1\cdots\mathcal{O}_n
\int_{\mathcal{I}^-_+}\Big(q_{\text{cl},c^A_1}+q_\Ps\Big)\right\rangle=0. 
\end{equation}
The irrelevance of the $s$-exact term is a consequence of 
\begin{equation} 
\bigg\langle\ \Big(sG_\Ps\Big)\ \mathcal{O}_1\cdots\mathcal{O}_n\bigg\rangle
=\bigg\langle s\Big(G_\Psi\mathcal{O}_1\cdots\mathcal{O}_n\Big)\bigg\rangle
\pm\bigg\langle G_\Psi\Big(\sum_{i=1}^n\mathcal{O}_1\cdots s\mathcal{O}_i
 \cdots\mathcal{O}_n\Big)\bigg\rangle=0,
\end{equation}
which holds true because the mean value of an $s$-exact term vanishes and    
$s\mathcal{O}_i=0$ since the $\mathcal{O}_i$'s are in the cohomology of the
BRST operator. 
Then, to obtain the consequence of the Ward identity \eqref{WI_q_gauge} on
the ghost number zero sector, one may differentiate once with respect to the
primary ghost $c^A_1$.\footnote{It is a common practice to get non-trivial
consequences on the ghost number zero sector from a ghost number one BRST
Ward identity.
For example, one proves the gauge independence and the unitarity properties
for the physical $\mathcal{S}$-matrix from the bulk BRST master equation of
a gauge theory.}
To establish a $\lambda^A$ dependent equation, we apply the ghost number $-1$
operator $\lambda^A(y)\frac{\delta }{\delta c^A_1(y)}$ to both sides of 
\eqref{WI_q_gauge}, that is
\begin{equation}\label{GN0_WI}
\left.\int_M d^D y\ \lambda^A(y)\frac{\delta\eqref{WI_q_gauge}}{\delta c^A_1(y)}
   \right\vert_{c^A_1=c^P_2=0} \,  ,
\end{equation}
where the (ghost number zero) classical infinitesimal gauge parameter
$\lambda^A(y)$ is in a formal one to one correspondence with the primary
ghost $c^A_1(y)$.

Because of \eqref{explict_gauge_dependence}, the only term in
$q_{\text{cl},c^A_1 }+q_\Ps$ which survives the operation \eqref{GN0_WI} is
the ghost number one BRST Noether charge $q_{\text{cl},c^A_1 }$ stemming from
the classical action.
Moreover, in order not to lose half of the information of \eqref{WI_q_gauge}
when applying \eqref{GN0_WI}, we need to antipodally identify the ghost field
components
$c^A_1\big\vert_{\mathcal{I}^+_-}=c^A_1\big\vert_{\mathcal{I}^-_+}$.
The standard antipodal matching condition $\lambda^A(z,\bar{z})
\big\vert_{\mathcal{I}^+_-}=\lambda^A(z,\bar{z})\big\vert_{\mathcal{I}^-_+}$
\cite{Strominger_BMS_scattering} of the large gauge symmetry parameters that
leads to non-vanishing charges \eqref{corner_charge} naturally follows. 

We can finally apply the LSZ reduction formula to \eqref{GN0_WI} to obtain 
\begin{equation}\label{charge_ward_identity}
\bra{\rm out}[Q_{\lambda^A},\mathcal{S}]\ket{\rm in}=0,
\end{equation}
with the notations $\left\langle\mathcal{O}_1\cdots\mathcal{O}_n
\right\rangle\big\vert_{\rm LSZ}=\bra{\rm out}\mathcal{S}\ket{\rm in}$
and $[Q_{\lambda^A},\mathcal{S}]
=Q^+_{\lambda^A}\mathcal{S}-\mathcal{S}Q^-_{\lambda^A}$, where
$Q^\pm_{\lambda^A}$ are given by \eqref{corner_charge} integrated over
$\mathcal{I}^+_-$ and $\mathcal{I}^-_+$, respectively.
All this provides a \textit{universal gauge independent derivation} of the
invariance of the $\mathcal{S}$-matrix under asymptotic symmetries.
The holographic Ward identity \eqref{charge_ward_identity} is a physical
consequence of the BRST Noether 1.5th Theorem~\ref{claim_BRST_1.5}.
The potential loop corrections to this Ward identity have been algebraically
classified in~\cite{Baulieu:2024oql}.
They correspond to loop corrections of the associated soft theorem
\cite{Bern:2014oka,He_2014,Sahoo:2018lxl,He:2017fsb,Donnay:2022hkf,
Agrawal:2023zea,Choi:2024ygx}.  

A last important remark is that we need to assume that the $ \ket{\rm in}$
and $ \ket{\rm out}$ states are in the same asymptotic symmetry frame of
the degenerate vacuum after taking the LSZ reduction formula.
This constitutes another matching condition and it is necessary
in order to derive soft theorems from \eqref{charge_ward_identity} by using
standard perturbative quantum field theory techniques.

\section{Symplectic structure}\label{Section_symplectic}

In the covariant phase space approach to asymptotic symmetries, the symplectic structure plays a central role: it provides the geometric framework in which conserved charges are defined, their algebra computed, and the action of asymptotic symmetries on the space of solutions to the equations of motion encoded. 
In most analyses,  this symplectic structure is derived from the classical ungauge fixed Lagrangian, yielding a degenerate presymplectic form whose kernel corresponds to the gauge directions.
One must then perform a quotient procedure to obtain a non-degenerate symplectic form on the reduced phase space.  
The transformations that survive this reduction are the physical asymptotic symmetries, and their associated charge algebra projectively represents the asymptotic symmetry algebra used in both the classical and quantum analysis.

The goal of this section is to revisit this construction from the viewpoint of the BRST formalism. 
Starting from the gauge fixed and BRST invariant Lagrangian \eqref{L_GF},  we will analyze the associated gauge fixed symplectic structure.
It directly yields a non-degenerate symplectic form on the BRST covariant phase space.  
This construction provides an equivalent,  yet conceptually cleaner, foundation for defining conserved charges and their algebra: it automatically implements the removal of gauge redundancies, avoids the need of an explicit quotient procedure, and takes care of the gauge dependence through the BRST covariance. 
In particular, the resulting charge algebra is gauge independent and honestly represents the asymptotic symmetry algebra, thereby offering a cohomologically sound justification of the standard covariant phase space construction. 

In this section, we repeatedly make use of trigraded local forms that are
part of the BRST covariant phase space \cite{Baulieu:2024oql}.
We recall their definition and basic properties in Appendix~\ref{Annexe_A}.
Since we start with the space of fields and a Lagrangian, the
framework is in the Lagrangian formalism, not in the Hamiltonian formalism,
except a closed two-form can only be defined upon choosing a Cauchy
hypersurface in the spacetime.

\subsection{Preliminaries}

Let us start by briefly recalling the classical picture without gauge-fixing in order to set-up the notations and highlight the challenges to be addressed.

We work with the space of classical fields~$\phi^i$. 
A symplectic structure can always be defined from the  local
symplectic potential $\theta^\mu$ of the gauge invariant Lagrangian $L$ appearing in \eqref{fundam_identities}, which is
a one-form on the classical field space.  
Indeed, if we consider its associated spacetime codimension-one form $\theta$,  
this leads to a field space two-form
\begin{equation}\label{Symplectic_2}
\O\equiv\int_{\varSigma}\delta\theta,
\end{equation}
which is defined for an arbitrary codimension-one hypersurface $\varSigma$
in the spacetime.
If non-degenerate, $\O$ equips a symplectic structure on the classical covariant
phase space.
We note that although the space of fields and the subspace of on-shell
fields are completely covariant, the symplectic form $\O$ depends on a
choice of slice $\varSigma$.
However, the spacetime manifold needs not be a global product
$\varSigma\times\mathbb{R}$, and the two-form $\O$ does not change under
continuous bulk deformation of $\varSigma$ or, more generally, when $\varSigma$
represents the same relative homology cycle in the spacetime.
If the deformation occurs at the \emph{corners} $\pa \varSigma$, then $\O$ shifts only if a so-called symplectic flux is present \cite{Freidel}.

From $\O$,  one can define charges as follows. 
A vector field $V$ on field space is said to admit an \emph{Hamiltonian charge} $Q_V$ if  the \emph{fundamental canonical relation}
\begin{equation}\label{integrable_charge}
I_V\O\ \hat{=}\ \delta Q_V
\end{equation}
holds. 
 This relation is the Lagrangian counterpart of how moment maps usually
appears in the canonical Hamiltonian formalism.
If there exists a non-vanishing vector field $V$ such that $Q_V = 0$, then the symplectic two-form is degenerate and $V$ generates a small gauge transformation, that is, one that acts trivially on the physical phase space.   
One must therefore quotient the field space by the group of such transformations in order to eliminate the redundancies. The reduced (quotiented) on-shell field space is the covariant phase space on which 
the right-hand side of \eqref{integrable_charge}  vanishes only for the
zero vector field $V=0$, and $\O$ becomes
invertible. 
The Poisson bracket between two Hamiltonian charges $Q_{V_1}$ and $Q_{V_2}$ is defined by inverting $\O$:
\begin{equation}\label{poisson_bracket}
  \{Q_{V_1},Q_{V_2}\} \equiv L_{V_2}Q_{V_1} \ \hat{=}\ I_{V_1}I_{V_2}\O,  
\end{equation}
and it satisfies
\begin{equation}\label{canonical_action}
I_{V_1}\delta\Ph^I\ \hat{=}\ \{Q_{V_1},\Ph^I\} .
\end{equation}
So the charge $Q_{V_1}$ canonically generates the symmetry associated with $V_1$ on the covariant phase space. 
This Lagrangian construction provides the analogue of the
Poisson brackets defined in the Hamiltonian formalism.

The charge algebra \eqref{poisson_bracket} then provides a projective representation, i.e., up to possible central extensions, of the symmetry algebra generated by the Lie bracket $[V_1,V_2]$ \cite{Brown:1986nw}. The reduced phase space can then be formally quantized by promoting the Poisson bracket of charges to a quantum commutator.

However, this standard procedure is not always directly applicable.  Its validity depends on the precise nature of the residual symmetry generators $V$ that survive the reduction.
In fact, when dealing with the vector field $V_\lambda$ associated with asymptotic symmetries, 
the Noether charge $Q_{V_\lambda} = Q_\lambda$ in \eqref{Noether_corner_charge_classical} is not always Hamiltonian. 
This means that when  computing $I_{V_\lambda} \O$, we do not obtain an exact   form on the covariant phase space.
In such cases, one obtains
\begin{equation}\label{non_integrable_charge}
I_{V_\lambda} \O \ \hat{=}\ \delta Q_\lambda + \mathcal{F}_\lambda,
\end{equation}
where $ \mathcal{F}_\lambda$ is a non-exact $1$-form on the field space typically supported on spacetime corners.  
This $\mathcal{F}_\lambda$, called the \emph{Noetherian flux},  is the obstruction to $Q_\lambda$ being Hamiltonian.
 
When the asymptotic symmetry parameter $\lambda$ is  field independent, 
that is $\delta \lambda =0$, 
and when the symplectic anomaly (to be precisely defined in the next section) vanishes,  the Noetherian flux reduces to the symplectic flux responsible for shifts of $\O$ when we deform the corners~$\pa \varSigma$. 
In situations where $\mathcal{F}_\lambda = 0$, the Noether charge is said to be \emph{integrable} and coincides with the Hamiltonian charge \eqref{integrable_charge}.
In this sense,   $\mathcal{F}_\lambda$ is the \emph{non-integrable} piece of the fundamental canonical relation.

Obstructions to integrability arise for instance in open systems (notably asymptotically flat gravity) or whenever the asymptotic symmetry parameter is field dependent (see the reviews \cite{Compere:2019qed,Ciambelli:2022vot} for examples). In these cases, one cannot rely on the naive Poisson bracket in~\eqref{poisson_bracket}. 
One way around this problem is to define a modified bracket which incorporates the non-integrable contributions and is engineered so that the symmetry action is still realized by
\eqref{canonical_action} but for this modified
bracket \cite{Barnich_Charge_algebra,Freidel}.
The construction of this bracket crucially relies on the split between the integrable and non-integrable part of \eqref{non_integrable_charge}.
An alternative strategy is to enlarge the phase space by introducing some corner or ``edge" degrees of freedom \cite{Donnelly:2016auv,Speranza:2017gxd,Freidel:2020xyx,Freidel:2020svx,
Freidel:2020ayo,Ball:2024hqe,Ball:2024xhf}.
In this enlarged phase space the corner obstruction $\mathcal{F}_\lambda$ is absorbed in an extended symplectic form 
with edge modes.
Then the Noether charges become Hamiltonian and the standard Poisson bracket \eqref{poisson_bracket} can be employed on the extended  phase space~\cite{Ciambelli_2022_Embeddings,
freidel2021canonical,Ciambelli:2021vnn,Adami:2024gdx}.

We now turn to the new contribution of this work and explain how the approaches reviewed above can be implemented using a gauge fixed symplectic structure on the BRST covariant phase space. 
Accordingly, we work on the BRST field space \eqref{BRST_multiplet}, which contains the classical fields together with ghosts, antighosts and auxiliary fields.  The gauge fixed symplectic structure is obtained from the local symplectic potential $\tilde{\theta}^\mu$ of the gauge fixed Lagrangian $\tilde{L}$ in \eqref{fundam_identities}. 
The gauge fixed symplectic two-form is  
\begin{equation}
\label{Symplectic_2_GF}
\tilde{\O} \equiv\int_{\varSigma}\delta\tilde{\theta},
\end{equation}
where $\tilde{\theta}$ is the spacetime codimension-one and field space one-form associated with $\tilde{\theta}^\mu$. The local symplectic two-form is $\tilde{\o}^\mu \equiv \delta \tilde{\theta}^\mu$.

Working in the gauge fixed setting, the original gauge parameters are replaced by ghost fields and the relevant symmetry is the BRST symmetry. To test whether $\tilde{\O}$ is degenerate one therefore computes its contraction along the ghost number one vector field 
$V_{\text{\tiny BRST}}$ defined in  \eqref{V_BRST_main}. 
This computation involves
the off-shell and local \emph{BRST fundamental canonical relation}
\begin{equation}\label{GF_fund_canon_1}
 I_{V_{\text{\tiny BRST}}}\tilde{\o}^\mu= \delta J^\mu_{\rm BRST}-Z^\mu+\pa_\nu Y^{\mu\nu} ,
\end{equation}
whose detailed derivation is given in Appendix~\ref{Annexe_A}. 
The quantities $Z^\mu$ and
$\pa_\nu Y^{\mu\nu}$ are related to those in \eqref{fundam_identities} by
\begin{align}\label{def_Z_Y}
\pa_\mu Z^\mu&=s(\tilde{E}_{\Ph^I}\delta\Ph^I),     \nn \\
\pa_\nu Y^{\mu\nu}&=s\tilde{\theta}^\mu+\delta K^\mu+Z^\mu. 
\end{align}
The corner quantity $Y^{\mu\nu}$, which is antisymmetric in $(\mu,\nu)$,
will be referred to as the \textit{BRST Noetherian flux}. 

An immediate and crucial consequence of \eqref{GF_fund_canon_1}, together with the Noether 1.5th Theorem~\ref{claim_BRST_1.5}, $J_{\rm BRST} \ \hat{=} \  s G_\Ps + d ( q_{\text{cl},c^A_1}+q_\Ps )$, is that the gauge fixed symplectic two-form $\tilde{\O}$ is generically non-degenerate and hence invertible on the BRST covariant phase space. 
Indeed,  even in theories that admit no asymptotic symmetries and no gauge-dependent boundary charges $q_\Ps$,  so that the surface integral
\begin{equation}
\label{BRST_charges}
Q_{V_{\text{\tiny BRST}}} \equiv \int_{\pa \varSigma} \big(  q_{\text{cl},c^A_1}+q_\Ps \big) 
\end{equation} 
vanishes,  the global on-shell version of \eqref{GF_fund_canon_1} contains at least the contribution
\begin{equation}
\label{delta_s_exact}
\delta  s \int_{\varSigma} G_\Ps \subset I_{V_{\text{\tiny BRST}}} \tilde{\O}  ,
\end{equation}
which does not vanish.\footnote{By the explicit expression of $G_\Ps$ in the proof of the Noether 1.5th theorem, this statement requires that the gauge-fixing fermion $\Ps$ depend on the gauge field.  
This is satisfied for any standard gauge choice.}

Since $\tilde{\O}$ is invertible,  one may attempt to construct a bracket for the BRST charges~\eqref{BRST_charges} by considering the double contraction $I_{V_{\text{\tiny BRST}}} I_{V_{\text{\tiny BRST}}}  \tilde{\O}$. 
The $(\delta s)$-exact contribution~\eqref{delta_s_exact} drops out under this double contraction because
\begin{equation}
I_{V_{\text{\tiny BRST}}} \delta s  \int_{\varSigma} G_\Ps  =  s^2  \int_{\varSigma} G_\Ps  =0 
\end{equation}
 by the definition \eqref{defining_s_main} and the nilpotency of $s$. Hence we obtain
 \begin{equation}
 \label{tentative_BRST_poisson_bracket}
 I_{V_{\text{\tiny BRST}}} I_{V_{\text{\tiny BRST}}}  \tilde{\O} \ \hat{=} \ - L_{V_{\text{\tiny BRST}}} Q_{V_{\text{\tiny BRST}}} +  I_{V_{\text{\tiny BRST}}} \int_{\varSigma} (d^{D{-}1}x)_\mu \big( -Z^\mu+\pa_\nu Y^{\mu\nu} \big) . 
 \end{equation}
The first term on the right has the same form as the Poisson bracket \eqref{poisson_bracket},  but at this stage we cannot assume that the second term, coming from the non-integrable part of~\eqref{GF_fund_canon_1},  vanishes on-shell. 
We will show in the next section that, whenever $Q_{V_{\text{\tiny BRST}}} \neq 0$,  this term indeed fails to vanish on-shell. 
Therefore,  in the gauge fixed BRST setting one cannot rely on the naive Poisson bracket.

The two strategies used in the ungauge fixed context to handle this non-integrability (described below eq.~\eqref{non_integrable_charge}) are based on the precise form of the non-integrable term $\mathcal{F}_\lambda$. 
Our aim is to extend the modified bracket approach to the gauge fixed BRST setting. Concretely,  we therefore need (i) an explicit expression for the BRST Noetherian flux~$Y$—the ghost number one, field space $1$-form and spacetime codimension-two form associated with~$Y^{\mu\nu}$ in \eqref{def_Z_Y}—in terms of the gauge-dependent charges $q=q_{\text{cl},c^A_1}+q_\Ps$ appearing in the BRST Noether current \eqref{Noether_1.5_form}, and (ii) the on-shell evaluation of $Z^\mu$.

\subsection{BRST Noetherian flux}\label{section_BRST_flux}

We start by analyzing the classical part $Y_{\rm cl}$ of $Y$, namely the part
that does not depend on the gauge fixing functional $\Ps$ but simply on the
gauge invariant Lagrangian $L$.
We use the tools of Appendix~\ref{Annexe_A} and introduce the BRST anomaly
operator to express $Y_{\rm cl}$ in terms of the classical corner Noether
charges $q_{\text{cl},c^A_1}$.

The analogue of $Y_{\rm cl}$ when working with gauge parameters $\lambda$ instead of ghosts is the quantity $\mathcal{F}_\lambda$ in \eqref{non_integrable_charge}. 
In this context,  it has been shown in \cite{Freidel}
that the most general expression for $\mathcal{F}_\lambda$ is obtained by introducing
the so-called \emph{anomaly operator}
\cite{Freidel,Hopfmuller:2018fni,Chandrasekaran:2020wwn,Odak:2022ndm}.
We now extend this construction to the case where ghost fields are present, thereby determining~$Y_{\rm cl}$.

Given that the BRST operator \eqref{BRST_trans} may describe both an internal
gauge symmetry with ghosts $\tilde{c}^a$ and a spacetime symmetry with vector
ghost $\xi$, we define 
\begin{equation}
\mathsf{C}\equiv c^A_1 \equiv ( \tilde{c}^a,\xi ) .
\end{equation}
If the theory under study has no spacetime symmetry, then $\C$ simply reduces
to $\tilde{c}^a$.
Here we keep $\tilde{c}^a$ and $\xi$ to stay as general as possible. 
In this case, as explained in Appendix~\ref{Annexe_A}, we can define the Lie
derivative in spacetime $\Lie_\xi=i_\xi d-di_\xi$ along the ghost vector field $\xi$ and
we know that the action of the BRST operator on the fields $\phi^i$ will
contain such a Lie derivative.
It is thus interesting to compare the action of $s$ with that of $\Lie_\xi$
by defining 
\begin{equation}\label{def_hat_s}
\hat{s}\equiv s-\Lie_\xi. 
\end{equation}
This operator isolates the part of the BRST transformation that does not coincide with the spacetime Lie derivative. 
Since $s$ contains the action of diffeomorphisms in field space (through $s=L_{V_{\text{\tiny BRST}}}$),  while $\Lie_\xi$ implements the geometric Lie derivative in spacetime,  their difference measures the departure from strict spacetime covariance.
By using the trigraded commutations rules of Table~\ref{table_2}, one has
\begin{equation}\label{hat_s_2}
\hat{s}^2=-\Lie_\phi 
\end{equation}
for $\phi\equiv s\xi-\demi\Lie_\xi\xi$.
The physical relevance of this $\phi$ is discussed in Appendix~\ref{Annexe_A}.

To extend the definition of the anomaly operator \cite{Freidel,Hopfmuller:2018fni,Chandrasekaran:2020wwn,Odak:2022ndm} to the BRST setting, 
we introduce the following construction. 
Any ghost number one field space scalar $X$ that is linear in the ghost $\C$ (from now on, we shall denote every such object by~$X_\C$) defines, by replacement $\mathsf{C} \to \delta\mathsf{C}$,  a ghost number one field space $1$-form: 
\begin{equation}
\label{def_X_deltaC}
X_{\delta\mathsf{C}}\equiv
\delta c^A_1 \, \frac{\pa X}{\pa c^A_1} + \pa_\mu (\delta c^A_1) \,  \frac{\pa X}{\pa \pa_\mu c^A_1} + \pa_\mu \pa_\nu (\delta c^A_1) \,  \frac{\pa X}{\pa \pa_\mu \pa_\nu c^A_1}   + \dotsc 
= X_{\mathsf{C}}\big\vert_{\mathsf{C} \to \delta\mathsf{C}} \, . 
\end{equation} 
The rationale for introducing $X_{\delta \C}$ is that ghosts are genuine dynamical fields on the BRST covariant phase space,  so their field space variation $\delta \C$ is non-zero, unlike the variation of a field-independent gauge parameter \(\lambda\) for which \(\delta\lambda=0\). 
These $\delta \C$-dependent contributions must therefore be tracked when taking \(\delta\)-variations that enter the derivation of the fundamental canonical relation, and thus of $Y_{\rm cl}$.

For example, let $V_{\text{\tiny BRST}}$  be the field space vector field \eqref{V_BRST_main} that generates the BRST transformations \eqref{BRST_trans}. 
When acting on a field space scalar \(f(\phi^i)\) depending only on the classical fields, we define 
\begin{equation}
\label{def_linear_V_C}
V_{\C} (f) \equiv V_{\text{\tiny BRST}} ( f ) = I_{V_{\text{\tiny BRST}}} \delta f = \int_M dx \ s\phi^i(x)\frac{\delta f(\phi^j)}{\delta\phi^i(x)}  .
\end{equation}
By \eqref{BRST_trans},  $V_{\C} (f)$ is a ghost number one field space scalar linear in $\C$,  so the definition~\eqref{def_X_deltaC} gives
\begin{equation}
V_{\delta \C} (f)  = \int_M dx \ \big[ s\phi^i(x) \big] \Big\vert_{\mathsf{C} \to \delta\mathsf{C}} \ 
\frac{\delta f(\phi^j)}{\delta\phi^i(x)}  .
\end{equation}
This formula defines $V_{\delta \C}$:  it is a commuting, ghost number one field space vector field valued in field space $1$-forms.

We can now define the \textit{BRST anomaly operator} as
\begin{equation}\label{def_BRST_anomaly_op}
\varDelta_\C\equiv\hat{s}+I_{V_{\delta\C}}=L_{V_\C}-\Lie_\xi+I_{V_{\delta\C}} .
\end{equation}
Its action is defined only on field space forms constructed from the classical fields~$\phi^i$.
Since in this section we are concerned only with the classical contribution to $Y$,   arising from the classical symplectic structure \eqref{Symplectic_2} that is independent of ghosts, antighosts and auxiliary fields, this is all we need.  
By making use of Table~\ref{table_2}, one gets the following graded
commutation relations 
\begin{align}\label{Delta_commutation}
\delta\varDelta_\C&=-\varDelta_\C\delta+\varDelta_{\delta\C}, & d\varDelta_\C
 &=-\varDelta_\C d,                                                    \nn \\
I_{V_\C}\varDelta_\C&=\varDelta_\C I_{V_\C}+I_{V_{s\C}},& s\varDelta_\C
 &=-\varDelta_\C s-\Lie_{s\xi}-L_{V_{\delta\C}}I_{V_\C},
\end{align} 
where $\varDelta_{\delta\C}\equiv L_{V_{\delta\C}}-\Lie_{\delta\xi}$.
On classical field space scalars, the action of the BRST anomaly operator reduces to that of \eqref{def_hat_s}: 
\begin{equation}
\varDelta_\C \phi^i = \hat{s} \phi^i ,
\end{equation}
and therefore captures the failure of spacetime covariance.

Let us now use the operator \eqref{def_BRST_anomaly_op} to express
$Y_{\rm cl}$ in terms of 
\begin{equation}
q_{\delta \C} = q_\C \big\vert_{\C\to\delta  \C}  \equiv q_{\text{cl}, c^A_1}\big\vert_{c^A_1\to\delta c^A_1} .
\end{equation}
In the following computations,  we must pay extra attention to the various signs appearing in the
computations due to the trigrading.
For instance if we work with spacetime differential forms $\phi^i$ rather
than spacetime multi-vector fields,  we use
$I_{V_{\delta\C}}\delta\phi^i
=-(s\phi^i)\big\vert_{\mathsf{C}\to\delta\mathsf{C}}$.

We start by noticing that the gauge invariant Lagrangian $L$,  seen as a
spacetime top-form, can be \textit{anomalous} in the sense that 
\begin{equation}
sL=dK_\C\quad\quad\text{and}\quad\quad\Lie_\xi L=-di_\xi L\ne sL
\quad\quad\Longrightarrow\quad\quad\hat{s}L\ne0. 
\end{equation}
This may happen when the BRST symmetry is associated with more than just the
diffeomorphism symmetry labeled by $\xi$, or when one imposes specific
boundary conditions by adding a boundary Lagrangian to $L$, for instance the
Gibbons-Hawking boundary term in general relativity,  which breaks spacetime covariance as it relies on introducing a preferred normal direction to the boundary. 
We then define the  anomaly of a Lagrangian  as 
\begin{equation}\label{anomaly_L}
\varDelta_\C L=\hat{s}L=da_\C\qquad\text{with}\qquad a_\C=K_\C+i_\xi L, 
\end{equation}
which leads to the relation
\begin{equation}\label{delta_anomaly}
\varDelta_{\delta\C}L=-da_{\delta\C}. 
\end{equation}

We need other identities coming from the definition of the equations of
motion $E=\delta L-d\theta$. 
In Appendix~\ref{Annexe_A}, it is shown that 
\begin{align}
I_{V_\C}E&=dC_\C, \nn \\
I_{V_\C}\theta-a_\C+i_\xi L&=I_{V_\C}\theta-K_\C=-C_\C+dq_\C,
\end{align}
from which we deduce 
\begin{align}\label{I_V_delE_theta_E}
I_{V_{\delta\C}}E&=-dC_{\delta\C},  \nn \\
 I_{V_{\delta \C}} \theta  &=  - C_{\delta \C} - d q_{\delta \C}  + K_{\delta \C} = - C_{\delta \C} - d q_{\delta \C} + a_{\delta \C} - i_{\delta \xi} L  . 
\end{align}
By using \eqref{def_Z_Y} and \eqref{I_V_delE_theta_E}, we can now compute
the anomaly of the equations of motion.
We get 
\begin{align}\label{anomaly_E}
\varDelta_\C E&=sE-\Lie_\xi E+I_{V_{\delta \C}}E
=d(Z_{\rm cl}+i_\xi E-C_{\delta\C}).
\end{align}
The anomaly of the local symplectic potential is obtained by using
\eqref{Delta_commutation}, \eqref{anomaly_L}, \eqref{delta_anomaly},
\eqref{anomaly_E} and by computing
\begin{align}\label{anomaly_theta_ungauge}
-d\delta a_\C
&=\delta\varDelta_\C L=-\varDelta_\C\delta L+\varDelta_{\delta\C}L
=-\varDelta_\C E+d(\varDelta_\C \theta-a_{\delta\C})                    \nn\\
\Longrightarrow\quad\varDelta_\C \theta&\equiv dA_\C-\delta a_\C
 +a_{\delta\C}+Z_{\rm cl}+i_\xi E-C_{\delta\C}.
\end{align}
This equation defines the \textit{symplectic anomaly} $A_\C$.

We can finally compute the Noetherian flux $dY_{\rm cl}$ defined in
\eqref{def_Z_Y}.
We find 
\begin{align}\label{explicit_BRST_noetherian_flux}
dY_{\rm cl}=&\ s\theta+\delta K_\C-Z_{\rm cl}            \nn\\
=&\ \varDelta_\C\theta+i_\xi d\theta-di_\xi\theta-I_{V_{\delta\C}}\theta
 +\delta K_\C-Z_{\rm cl}                                 \nn\\
=&\ dA_\C-\delta a_\C+a_{\delta\C}+Z_{\rm cl}+i_\xi E-C_{\delta\C}+i_\xi
 \delta L-i_\xi E-d i_\xi\theta+C_{\delta\C}+dq_{\delta\C}-a_{\delta\C}  \nn\\
&+i_{\delta\xi}L+\delta(a_\C-i_\xi L)-Z_{\rm cl}         \nn\\
=& \ d\big(q_{\delta\C}-i_\xi\theta+A_\C\big).
\end{align}
This expression of the Noetherian flux is in agreement with that of
\cite{Freidel} modulo flips of signs because of the different statistics due
to the ghost number.
Notice that we didn't have to assume that the equations of motion were anomaly
free, namely that $\varDelta_\C E=0$, to derive this result.
Another important thing to notice is that the BRST Noetherian flux does not
depend on the anomaly of the Lagrangian but only on the symplectic anomaly.
The expression \eqref{explicit_BRST_noetherian_flux} also shows that in the
BRST case, since $\delta \C$ never vanishes, the local fundamental canonical
relation \eqref{GF_fund_canon_1} always has a non-integrable part. 
Hence as soon as $Q_{V_{\text{\tiny BRST}}} \neq 0$,  namely whenever asymptotic symmetries are present, the second term in the right-hand side of \eqref{tentative_BRST_poisson_bracket} does not vanish and the Poisson bracket cannot be used.

\subsection{Gauge fixed fundamental canonical relation}
\label{section_gauge_fixed_fund}

We now turn to the determination of the gauge dependence of the non-integrable
part $-Z^\mu+\pa_\nu Y^{\mu\nu}$ of the fundamental canonical relation
\eqref{GF_fund_canon_1} on-shell.
We come back to spacetime multi-vector fields for the following computations
because it is easier for keeping track of the integrations by parts. 

The computation of $Z^\mu$ has already been done in \eqref{computation_Z}.
Indeed, by using the constraints $G_{\Ph^I}=0$, we obtain
\begin{align}\label{Z_on-shell}
Z^\mu &=-(-1)^{\epsilon(\tilde{E}_{\phi^i})}\tilde{E}_{\phi^i}
 \left(\frac{\pa f^i_B}{\pa\pa_\mu\phi^j}\delta\phi^j c^B_1
 +f^{\mu i}_B\delta c^B_1\right)
 -(-1)^{\epsilon(\tilde{E}_{c^A_1})}\tilde{E}_{c^A_1}
 \left(M^{\mu A}_P\delta c^P_2-M^{\mu A}_{BC}c^B_1\delta c^C_1\right)  \nn \\
&\quad+(-1)^{\epsilon(\tilde{E}_{c^P_2})}\tilde{E}_{c^P_2}
 \varGamma^{\mu P}_{BQ}c^B_1\delta c^Q_2\quad\Longrightarrow\quad Z^\mu
 \ \hat{=}\ 0.
\end{align}
The fact that $Z^\mu$ vanishes on-shell was conjectured in
\cite{Baulieu:2024oql}, it is now proven by \eqref{Z_on-shell}. 
Notice that this is in agreement with the ungauge fixed case, where the on-shell
vanishing of $C_{\delta \C} = Z_{\rm cl} + d(\dotsc)$ is implied by Noether's second theorem.  

Getting the gauge dependence $\pa_\nu Y^{\mu\nu}_{\rm gauge}$ of
$\pa_\nu Y^{\mu\nu}$ is a bit more involved.
A useful thing to remember when performing this computation is the expression
$\pa_\nu Y^{\mu\nu}_{\rm cl}$ of the previous section. 
Guided by this expression, we would like to express
$\pa_\nu Y^{\mu\nu}_{\rm gauge}$ in terms of $q_\Ps\big\vert_{\C=\delta\C}$,
where $q_\Ps$ are the gauge charges defined in
\eqref{explict_gauge_dependence}.
This guess will be helpful when performing the various integrations by parts
appearing in the computation of $\pa_\nu Y^{\mu\nu}_{\rm gauge}$.

{}From \eqref{fundam_identities} we know that $K^\mu_{\rm gauge}=0$.
So by using \eqref{theta_gauge} and \eqref{Z_on-shell}, we get 
\begin{align}\label{Y_gauge}
\pa_\nu Y^{\mu\nu}_{\rm gauge}&=s\theta^\mu_{\rm gauge}+Z^\mu_{\rm gauge} \nn\\
&=(-1)^{\epsilon(F_{\phi^i})}\Big(s F_{\phi^i}+\tilde{E}_{\phi^i}-E_{\phi^i} 
 \Big)
\Big(\frac{\pa f^i_B}{\pa\pa_\mu\phi^j}\delta\phi^j c^B_1+f^{\mu i}_B\delta
 c^B_1\Big)                                                               \nn\\
&\quad-(-1)^{\epsilon(F_{c^A_1})}\Big(sF_{c^A_1}+\tilde{E}_{c^A_1}\Big)
\Big(M^{\mu A}_{BC}c^B_1\delta c^C_1-M^{\mu A}_P\delta c^P_2\Big)         \nn\\
&\quad-(-1)^{\epsilon(F_{c^P_2})}\Big(sF_{c^P_2}+\tilde{E}_{c^P_2}\Big)
 \varGamma^{\mu P}_{BQ}c^B_1\delta c^Q_2                                  \nn\\
&\quad+F_{\phi^i}s\Big(\frac{\pa f^i_B}{\pa\pa_\mu\phi^j}\delta\phi^j c^B_1
 +f^{\mu i}_B\delta c^B_1\Big)-F_{c^A_1}s
 \Big(M^{\mu A}_{BC}c^B_1\delta c^C_1-M^{\mu A}_P\delta c^P_2\Big)        \nn\\
&\quad-F_{c^P_2}s\Big(\varGamma^{\mu P}_{BQ}c^B_1\delta c^Q_2\Big). 
\end{align}
We will divide the computation of this boundary term in three main steps,
each corresponding to the dependence of \eqref{Y_gauge} in \textit{(i)}
$F_{\phi^i}$, \textit{(ii)} $F_{c^A_1}$ and \textit{(iii)} $F_{c^P_2}$.
Given that this computation must be valid off-shell, we use the definition of
the equations of motion in~\eqref{theta_gauge} to reexpress
$sF_{\Ph^I}+\tilde{E}_{\Ph^I}$ in \eqref{Y_gauge}.
We must also remember that our parametrization for the BRST transformations
\eqref{BRST_trans} is such that $\frac{\pa f^i_B}{\pa\pa_\mu\phi^j}$ is field
independent so $s\Big(\frac{\pa f^i_B}{\pa\pa_\mu\phi^j}\Big)=0$.\\

\noindent\textit{(i)} The $F_{\phi^i}$ dependent terms in \eqref{Y_gauge} are:
\begin{align}\label{Y_F_phi}
\pa_\nu Y^{\mu\nu}_{\rm gauge} &\supset \bigg( -    F_{\phi^k} \Big( \frac{\pa f^k_A}{\pa \phi^i} c^A_1  + \frac{\pa f^{\nu k}_A}{\pa \phi^i} \pa_\nu c^A_1 \Big)  +   \pa_\nu \Big( F_{\phi^k} \frac{\pa f^k_A}{\pa \pa_\nu \phi^i} c^A_1 \Big)  \bigg) \bigg( \frac{\pa f^i_B}{\pa \pa_\mu \phi^j} \delta \phi^j c^B_1 + f^{\mu i}_B \delta c^B_1    \bigg) 
\nn \\
&\quad + \bigg( F_{\phi^i} f^i_A - \pa_\nu \Big( F_{\phi^i} f^{\nu i}_A \Big) \bigg)  \bigg( M^{\mu A}_{BC} c^B_1 \delta c^C_1 - M^{\mu A}_P \delta c^P_2 \bigg)
- F_{\phi^i} \frac{\pa f^i_B}{\pa \pa_\mu \phi^j} \bigg(  \delta \Big( f^j_A   c^A_1  
\nn \\
&\quad  + f^{\nu j}_A   \pa_\nu c^A_1 \Big) c^B_1  + \delta \phi^j \Big(  \gamma^B_{\ P}  c^P_2 + M^{\nu B}_P   \pa_\nu c^P_2 +   \gamma^B_{\ AC}   c^A_1 c^C_1 + M^{\nu B}_{A C}  c^A_1 \pa_\nu c^C_1  \Big)   \bigg)
\nn \\
&\quad + F_{\phi^i}   \bigg( \frac{\pa f^{\mu i}_B}{\pa \phi^j} s \phi^j \delta c^B_1  - f^{\mu i}_B \delta \Big(  \gamma^B_{\ P}  c^P_2 + M^{\nu B}_P   \pa_\nu c^P_2 +   \gamma^B_{\ AC}   c^A_1 c^C_1 + M^{\nu B}_{A C}  c^A_1 \pa_\nu c^C_1   \Big)  \bigg).
\end{align} 
We can now group these terms by their $1$-form in field space dependence.
By using \eqref{27}, \eqref{28} and one integration by parts, the
$\delta c^P_2$ terms of the $1$-form regroup into a boundary term, which is
\begin{equation}\label{Y_1}
\pa_\nu Y^{\mu\nu}_{\rm gauge}\supset\pa_\nu
 \Big(F_{\phi^i}f^{\nu i}_AM^{\mu A}_P\delta c^P_2\Big).
\end{equation}
As it should, this boundary term is antisymmetric in $(\mu,\nu)$ by
\eqref{27}. 

The $\delta c^A_1$ terms in \eqref{Y_F_phi} form a boundary term by \eqref{210},  
\eqref{211}, \eqref{212} and two integrations by parts, namely 
\begin{equation}\label{Y_2}
\pa_\nu Y^{\mu\nu}_{\rm gauge}\supset\pa_\nu\bigg(F_{\phi^i}
 \Big(\frac{\pa f^i_B}{\pa\pa_\nu\phi^j}f^{\mu j}_A
 -f^{\nu i}_C M^{\mu C}_{BA}\Big)c^B_1\delta c^A_1\bigg).
\end{equation}
It is indeed antisymmetric in $(\mu,\nu)$ by \eqref{211}. 

For the $\delta \phi^j$ terms of the $1$-form, we need additional constraints.
They come from imposing $\frac{\pa(s^2\phi^i)}{\pa\pa_\mu\phi^j}=0$.
In fact, since the consequences of $s^2 \phi^i=0$ are obtained by expanding in
a polynomial basis of ghost fields and their derivatives, as in 
\eqref{s2cphi}, imposing $\frac{\pa(s^2\phi^i)}{\pa\pa_\mu\phi^j}=0$ leads us
to new constraints that are basically the derivative with respect to
$\pa_\mu\phi^j$ of the constraints \eqref{26}-\eqref{212}.
The fact that the parametrizing functions of \eqref{BRST_trans} are polynomial
in the fields is crucial here to be able to use identities of the type 
\begin{equation}
\frac{\pa(\pa_\nu\gamma^A_{\ BC})}{\pa\pa_\mu\phi^j}
=\delta^\mu_\nu\frac{\pa\gamma^A_{\ BC}}{\pa\phi^j}.
\end{equation}
We can now use the new constraints coming from \eqref{26}, \eqref{28},
\eqref{29}, \eqref{210} and one integration by parts in \eqref{Y_F_phi} to
show that the $\delta\phi^j$ terms of the $1$-form indeed reduce to a
boundary term, that is 
\begin{equation}\label{Y_3}
\pa_\nu Y^{\mu\nu}_{\rm gauge}
\supset\pa_\nu\bigg(F_{\phi^i}\frac{\pa f^i_A}{\pa\pa_\nu\phi^j}
 \frac{\pa f^j_B}{\pa\pa_\mu\phi^k}c^A_1\delta\phi^kc^B_1\bigg)=0. 
\end{equation}
This boundary term  vanishes by a new constraint that comes from acting with
$\frac{\pa}{\pa(\pa_\nu\pa_\mu\phi^k)}$ on the constraint \eqref{29}.
It is actually not a surprise that this boundary term does not contribute
since there is no way to prove its antisymmetry in $(\mu,\nu)$.\\

\noindent\textit{(ii)} The $F_{c^A_1}$ dependent terms in \eqref{Y_gauge} are:
\begin{align}\label{Y_F_c1}
\pa_\nu Y^{\mu\nu}_{\rm gauge} &\supset   - F_{c^A_1} \frac{\pa  }{\pa \phi^i} 
\big( sc^A_1 \big)
 \Big(  \frac{\pa f^i_B}{\pa \pa_\mu \phi^j} \delta \phi^j c^B_1 + f^{\mu i}_B \delta c^B_1  \Big)                      \nn \\
&\quad -  \Big( F_{c^D_1} \big( \gamma^D_{\ \{AE\}}   c^E_1  + M^{\nu D}_{AE} \pa_\nu c^E_1  \big)  + \pa_\nu \big( F_{c^D_1} M^{\nu D}_{EA} c^E_1 \big)  \Big)  \Big( M^{\mu A}_{BC} c^B_1 \delta c^C_1 - M^{\mu A}_P \delta c^P_2  \Big)    \nn \\
&\quad+\Big( F_{c^D_1} \gamma^D_{\ P} - \pa_\nu \big(   F_{c^D_1} M^{\nu D}_P  \big)   \Big)  \varGamma^{\mu P}_{BQ} c^B_1 \delta c^Q_2
 - F_{c^A_1} s \Big( M^{\mu A}_{BC} c^B_1 \delta c^C_1 - M^{\mu A}_Q \delta c^Q_2 \Big) .
\end{align}
By using \eqref{215}, \eqref{216}, \eqref{217} and two integrations by parts,
the $\delta c^P_2$ terms of the $1$-form regroup into a boundary term, which
is
\begin{equation}\label{Y_4}
\pa_\nu Y^{\mu\nu}_{\rm gauge}\supset\pa_\nu
 \bigg(F_{c^D_1}\Big(M^{\nu D}_{EA}M^{\mu A}_Q
 -M^{\nu D}_P\varGamma^{\mu P}_{EQ}\Big)c^E_1\delta c^Q_2\bigg).
\end{equation}
As it should, this boundary term is antisymmetric in $(\mu,\nu)$ by
\eqref{215}.

The $\delta c^A_1$ terms of the $1$-form become a boundary term by
\eqref{214}, \eqref{217}, \eqref{219}, \eqref{220}, \eqref{221} and one
integration by parts, namely 
\begin{equation}\label{Y_5}
\pa_\nu Y^{\mu\nu}_{\rm gauge}\supset-\pa_\nu
 \Big(F_{c^D_1}M^{\nu D}_{EA}M^{\mu A}_{BC}c^E_1c^B_1\delta c^C_1\Big).
\end{equation}
This term either vanishes or is antisymmetric in $(\mu,\nu)$ by \eqref{220}. 

For the $\delta \phi^j$ terms of the $1$-form, we find that they all vanish
by new constraints that come from acting with $\frac{\pa}{\pa(\pa_\mu\phi^j)}$
on the constraints \eqref{213}, \eqref{216}, \eqref{218} and \eqref{219}.\\

\noindent\textit{(iii)}
The $F_{c^P_2}$ dependent terms in \eqref{Y_gauge} are:
\begin{align}
\label{Y_F_c2}
\pa_\nu Y^{\mu\nu}_{\rm gauge}
&\supset-F_{c^P_2}\frac{\pa}{\pa\phi^i}\big(sc^P_2\big)
\Big(\frac{\pa f^i_B}{\pa\pa_\mu\phi^j}\delta\phi^jc^B_1
 +f^{\mu i}_B\delta c^B_1\Big)                                        \nn \\
&\quad+F_{c^P_2}\Big(\beta^P_{\ \{ACD\}}c^C_1c^D_1+\varGamma^P_{\ QA}c^Q_2
 +\varGamma^{\nu P}_{AQ}\pa_\nu c^Q_2\Big)\Big(M^{\mu A}_{EF}c^E_1\delta c^F_1
 -M^{\mu A}_R\delta c^R_2\Big)                                        \nn \\
&\quad+\Big(F_{c^P_2}\varGamma^P_{\ QC}c^C_1
 -\pa_\nu\big(F_{c^P_2}\varGamma^{\nu P}_{CQ}c^C_1\big)\Big)
  \varGamma^{\mu Q}_{DR}c^D_1\delta c^R_2
 -F_{c^P_2}s\Big(\varGamma^{\mu P}_{CQ}c^C_1\delta c^Q_2\Big) .   
\end{align}
The $\delta c^P_2$ dependent terms can be grouped into the boundary term 
\begin{equation}\label{Y_6}
\pa_\nu\Big(F_{c^P_2}\varGamma^{\nu P}_{CQ}\varGamma^{\mu Q}_{DR}c^C_1c^D_1
 \delta c^R_2\Big)\subset\pa_\nu Y^{\mu\nu}_{\rm gauge}
\end{equation}
by doing one integration by parts and making use of \eqref{223}, \eqref{224},
\eqref{226}, \eqref{227} and~\eqref{229}.
The boundary term \eqref{Y_6} again either vanishes or is antisymmetric in
$(\mu,\nu)$ by~\eqref{226}.

The $\delta c^A_1$ dependent terms can be shown to vanish by using
\eqref{227}, \eqref{228} and \eqref{231}. 
It also happens to be the case for the $\delta\phi^j$ ones by the constraints
obtained from acting with $\frac{\pa}{\pa(\pa_\mu\phi^j )}$ on \eqref{225},
\eqref{229} and \eqref{230}.
This concludes the computation of $\pa_\nu Y^{\mu\nu}_{\rm gauge}$. 

Given the explicit form \eqref{explict_gauge_dependence} of the gauge charges,
which can be written as  
\begin{equation}
q^{\mu\nu}_\Ps=q^{\mu\nu}_{\Ps, c^A_1}+q^{\mu\nu}_{\Ps,c^P_2},
\end{equation}
where $q_{\Ps,c^A_1}$ does not depend on $c^P_2$ and $q_{\Ps,c^P_2}$ is linear
in $c^P_2$, we can use the fact that
$\epsilon(F_{\phi^i})=\epsilon(F_{c^A_1})+1=\epsilon(F_{c^P_2})$ and the
results \eqref{Y_1}, \eqref{Y_2}, \eqref{Y_4}, \eqref{Y_5} and \eqref{Y_6}
to write 
\begin{align}
\pa_\nu Y^{\mu\nu}_{\rm gauge}&=(-1)^{\epsilon(F_{\phi^i})}\pa_\nu
 \Big(q^{\mu\nu}_{\Ps,\delta c^A_1}-\demi q^{\mu\nu}_{\Ps,\delta c^P_2}\Big).
\end{align}
This is close from our initial guess but with the very important differences
that there is no gauge
dependent equivalent of the symplectic anomaly $A_\C$ and that $i_\xi\theta_{\rm gauge}$ does not contribute.

By combining all the results of this section and of Section~\ref{section_1.5},
the gauge fixed fundamental canonical relation \eqref{GF_fund_canon_1} takes
the following form 
\begin{multline}\label{GF_fund_canon}
I_{V_{\text{\tiny BRST}}} \tilde{\o}^\mu \ \hat{=}\ \delta sG^\mu_\Ps
 +\pa_\nu\Big(-\delta q^{\mu\nu}_{\text{cl},c^A_1}
 +q^{\mu\nu}_{\text{cl},\delta c^A_1}-\big(i_\xi\theta\big)^{\mu\nu}
 +\big( A_{c^A_1}\big)^{\mu\nu}\Big)                                  \\
 +\pa_\nu\bigg(\delta\Big(q^{\mu\nu}_{\Ps,c^A_1}+q^{\mu\nu}_{\Ps,c^P_2}\Big)
 +(-1)^{\epsilon(F_{\phi^i})}\Big(q^{\mu\nu}_{\Ps,\delta c^A_1}
  -\demi q^{\mu\nu}_{\Ps,\delta c^P_2}\Big)\bigg) .
\end{multline}
This determines the explicit gauge dependence of the non-integrable part of
the fundamental canonical relation. 
Only the classical part of the symplectic flux $\int_{\pa \varSigma} i_\xi \tilde{\theta}$ contributes to the global version of this relation.   Therefore the gauge-fixing does not induce shifts in $\tilde{\O}$ when the boundary $\pa \varSigma$ of the Cauchy slice is deformed.
Note that \eqref{GF_fund_canon} slightly differs from the predictions made in~\cite{Baulieu:2024oql}, in
which the gauge charges  $q^{\mu\nu}_{\Ps}$ were not present. 
To reach this result, every single constraints coming from the nilpotency of
the BRST operator have been used, except for~\eqref{222}, as well as all
equations of motion \eqref{theta_gauge} for $\phi^j$, $c^A_1$ and~$c^P_2$. 

\subsection{Charge bracket}
\label{section_charge_bracket}

As explained in Section~\ref{Holographic_WI}, the global Noether charges 
\begin{equation}\label{Noether_charges_asg}
Q_\C\equiv Q_{\text{cl},c^A_1}
=\int_{\pa\varSigma}  q_{\text{cl},c^A_1}
\end{equation}
are non-vanishing only when the ghosts $c^A_1$ are valued in the asymptotic
symmetry algebra~$\mathfrak{asg}$.  
We refer to these particular ghost components as
\textit{large ghosts}\footnote{Here we use the word ``large" in the sense of
large gauge symmetry as explained in the Introduction
above~\eqref{Noether_corner_charge_classical}.}
and we call them~$\C$, with $q_\C \equiv q_{\text{cl}, \C}$. 
The structure of the asymptotic symmetry algebra $\mathfrak{asg}$ is then
completely encoded in the nilpotent action of the large BRST operator $s\equiv L_{V_\C}$.\footnote{Note that the action of this $V_\C$ is no longer restricted to classical fields as in \eqref{def_linear_V_C}, nor is it linear in~$\C$.  It is simply the full BRST vector field defined in \eqref{V_BRST_main}.  The subscript $\C$ only serves to emphasize that this is the large BRST operator that enters the definition \eqref{V_BRST_main} for this vector field.}
Indeed,  the fact that the algebra closes
is encoded in 
\begin{align}\label{asg_structure}
s\C \in \mathfrak{asg}    \, ,
\end{align}
which means that $ Q_{s\C} \neq 0$, 
while the Jacobi identity is simply 
\begin{align}\label{asg_structure_2}
s^2\C =0  \, .  
\end{align}
Such large ghosts and  BRST operator have been constructed for the
extended BMS4 symmetry, which is the asymptotic symmetry of asymptotically
flat gravity in four dimensions, in \cite{Baulieu_Tom_BMS}.
They satisfy  \eqref{asg_structure} and \eqref{asg_structure_2} by construction.\footnote{
When working in the framework of infinitesimal gauge parameters, the asymptotic symmetry algebra can be represented using modified brackets that account for the possible dependence of the asymptotic symmetry parameters on the fields.  As seen in~\eqref{asg_structure}, working with ghosts avoids these subtleties, making the BRST approach more convenient.}

To make the requirement  \eqref{asg_structure} more explicit and to establish the consistency and gauge independence of the classical charge bracket construction \cite{Freidel},  we now show that one can define an algebra of ghost-dependent Noether charges $Q_\C$ \eqref{Noether_charges_asg} that faithfully represents the asymptotic symmetry algebra \eqref{asg_structure}-\eqref{asg_structure_2} for any gauge fixing~$\Ps$.

The charge bracket for the Noether charges $Q_\C$ has to be constructed out of the symplectic $2$-form $\tilde{\O}$
\eqref{Symplectic_2_GF}, which is the building block of the Poisson bracket
\eqref{poisson_bracket} in the integrable case. 
When these  charges are non-integrable, as it is the case in
\eqref{GF_fund_canon}, we start with the Barnich--Troessaert bracket
$\{Q_\C,Q_\C\}_{\rm BT}\equiv I_{V_\C}I_{V_\C}\tilde{\O}$ \cite{Barnich_Charge_algebra}.
By integrating~\eqref{GF_fund_canon} over $\varSigma$, we thus get 
\begin{multline}\label{generic_Q,Q}
\{Q_\C,Q_\C\}_{\rm BT} \ \hat{=}-sQ_\C+ Q_{s\C}
  -\int_{\pa\varSigma}(i_\xi I_{V_\C}\theta-I_{V_\C}A_\C)           \\
 +sQ_{\Ps,c^A_1}+(-1)^{\epsilon(F_{\phi^i})}Q_{\Ps,sc^A_1}
 +sQ_{\Ps,c^P_2}-\demi(-1)^{\epsilon(F_{\phi^i})}Q_{\Ps,sc^P_2},
\end{multline}
where we have used $I_V \delta s G^\mu_\Ps = s^2 G^\mu_\Ps = 0$ and defined
\begin{equation}\label{global_gauge_charges}
Q_\Ps=Q_{\Ps,c^A_1}+Q_{\Ps,c^P_2}\equiv\int_{\pa\varSigma}(d^{D{-}2}x)_{\mu\nu} \ 
 q^{\mu\nu}_{\Ps,c^A_1}+\int_{\pa\varSigma} (d^{D{-}2}x)_{\mu\nu} \ 
 q^{\mu\nu}_{\Ps,c^P_2}.
\end{equation}

A first thing to notice with the bracket \eqref{generic_Q,Q} is that it
depends on the gauge fixing functional $\Ps$.
We certainly do not want this since it must provide a representation of the
asymptotic symmetry algebra which is gauge independent.
This bracket should also define the canonical action of the charges $Q_\C$ on the
solution space as in \eqref{canonical_action}.
Since this is the action of asymptotic symmetries on solution space, which
are non-trivial physical symmetries, it cannot depend on the gauge dependent
charges \eqref{global_gauge_charges}.
To circumvent these problems, we simply impose restrictive boundary conditions
on the antighosts $\bar{c}_I$ and on the auxiliary fields $b_I$ that
are compatible with the BRST transformations \eqref{BRST_trans},
\begin{align}\label{bc_antighosts}
\bar{c}_I\Big\vert_{\pa\varSigma}=0\quad\quad\text{and}\quad\quad
b_I\Big\vert_{\pa\varSigma}=0.
\end{align}
Then by using \eqref{explict_gauge_dependence}, we know that $Q_\Ps$ is a
sum of linear terms in $F_\phi$, $F_{c^A_1}$ and $F_{c^P_2}$.
But by definition \eqref{fundam_identities}, these $F$'s have ghost number
$-1$, $-2$ and $-3$, respectively, which means that they must depend on
$\bar{c}_I$.
So by making use of the boundary conditions \eqref{bc_antighosts}, we have 
\begin{equation}\label{condition_gauge_charges}
Q_\Ps=0.
\end{equation}
 
Physically, the boundary conditions \eqref{bc_antighosts} mean that the gauge
has been fixed everywhere except on the corners $\pa \varSigma$.
This is the only place where we can do that, because it is precisely here that
the gauge symmetry becomes ``large" or physical. 
All this was possible because the whole gauge dependence of the bracket
concentrates on the corners,  thanks to the BRST Noether 1.5th Theorem~\ref{claim_BRST_1.5} and the property \eqref{Z_on-shell}. 

We are left with the gauge independent bracket 
\begin{equation}
\{Q_\C,Q_\C\}_{\rm BT}=I_{V_\C}I_{V_\C}\tilde{\O}\ \hat{=}-sQ_\C+Q_{s\C}
 -\int_{\pa\varSigma}(i_\xi I_{V_\C}\theta-I_{V_\C}A_\C),
\end{equation}
which a priori does not represent the asymptotic symmetry algebra yet.
To this end, we must define a charge bracket $\{ \cdot , \cdot \}_L$ that represents  \eqref{asg_structure}-\eqref{asg_structure_2}, namely one that satisfies
\begin{align}\label{properties_bracket}
{\{Q_\C,Q_\C\}}_L\ \hat{=}\ Q_{s\C}\quad\quad \text{and} \quad\quad
{\big\{Q_\C,{\{Q_\C,Q_\C\}}_L\big\}}_L\ \hat{=}\ Q_{s^2\C}=0. 
\end{align}
Since the asymptotic symmetry algebra is generated by $\C$, 
the first equality implies the second. Therefore  we focus only on the first one. 
The reason for the subscript $L$ is that every single quantity that will
appear in the definition of this bracket can be derived from the gauge
invariant Lagrangian $L$.
A further requirement is that this bracket must be invariant under boundary
shifts of the Lagrangian $L\longrightarrow L'=L+d\,l$.
Indeed, there is an ambiguity in the split between integrable and
non-integrable parts of the fundamental canonical relation
\eqref{GF_fund_canon} that comes from the ambiguity in the definition of
the local symplectic potential $\delta L'=E'+d\theta'$ with $E'=E$ and 
\begin{align}
\theta'&=\theta-\delta l+d\nu. 
\end{align}
The requirement that the bracket \eqref{properties_bracket} is insensitive to this
ambiguity means that the bracket before and after the shift should satisfy the algebraic equality
\begin{equation}\label{boundary_shift_independence}
{\{Q_\C,Q_\C\}}_L-Q_{s\C}={\{Q'_\C,Q'_\C\}}_{L'}-Q'_{s\C}.  
\end{equation} 

By using the various results of Section~\ref{section_BRST_flux}, we claim
that a bracket satisfying \eqref{properties_bracket} and
\eqref{boundary_shift_independence} is given by  
\begin{align}\label{charge_bracket_local}
{\{Q_\C,Q_\C\}}_L&\equiv\demi I_{V_\C}I_{V_\C}\tilde{\O}
 +\int_{\varSigma}d\Big(\Lie_\xi q_\C-I_{V_\C}A_\C+\kappa-\demi i_\xi i_\xi L\Big),
\end{align}
where $\kappa$ is defined as
\begin{equation}
d\kappa\equiv\varDelta_\C a_\C-i_\phi L. 
\end{equation}
The definition of $\kappa$ follows from \eqref{hat_s_2}, \eqref{anomaly_L}
and the algebraic Poincar\'e lemma. 

Let us prove that \eqref{charge_bracket_local} has indeed the expected
features \eqref{properties_bracket}-\eqref{boundary_shift_independence} of
a bracket and represents the asymptotic symmetry algebra
\eqref{asg_structure}-\eqref{asg_structure_2} without central extension.
By using the results of Section~\ref{section_BRST_flux} and omitting the
$q_\Ps$ dependence of \eqref{GF_fund_canon} which won't contribute to the
global bracket because of \eqref{condition_gauge_charges}, we compute 
\begin{align}\label{computation_I_V_I_V_o}
\demi I_{V_\C} I_{V_\C} \tilde{\o} \ \hat{=}& \ \demi \Big(  s d q_\C + s I_{V_\C} \theta - I_{V_\C} Z_{\rm cl} + s K_\C  \Big)
\nn \\
=& \ I_{V_\C} s \theta + \demi \Big( s C_\C - I_{V_\C} Z_{\rm cl} \Big)
\nn \\
=& \ I_{V_\C} \varDelta_\C \theta + I_{V_\C} \Lie_\xi \theta - I_{V_\C} I_{V_{\delta \C}} \theta + \demi \Big( s C_\C - I_{V_\C} Z_{\rm cl} \Big)
\nn \\
=& \ I_{V_\C} \Big( d A_\C - \delta a_\C + a_{\delta \C} + Z_{\rm cl} + i_\xi E + I_{V_{\delta \C}} \theta + d q_{\delta \C} - a_{\delta \C} + i_{\delta \xi} L   \Big) - I_{V_\C} I_{V_{\delta \C}} \theta 
\nn \\
&+ \Lie_\xi d q_\C + \Lie_\xi a_\C - \Lie_\xi C_\C - \Lie_\xi i_\xi L  + \demi \Big( s C_\C - I_{V_\C} Z_{\rm cl} \Big) 
\nn \\
=& \ \demi \Big(  s C_\C + I_{V_\C} Z_{\rm cl}   \Big) - d \Big(   \Lie_\xi q_\C - I_{V_\C} A_\C  + \kappa	- \demi i_\xi i_\xi L - i_\xi C_\C  \Big) + d q_{s \C}.
\end{align}

We then use \eqref{classical_J} to write
$C^\mu_\C=E_{\phi^i}f^{\mu i}_Ac^A_1$ and the classical expression of
$Z^\mu_{\rm cl}$ given by \eqref{Z_on-shell} to compute\footnote{The flip of
sign in front of $I_{V_\C}Z^\mu_{\rm cl}$ comes from the difference between
\eqref{GF_fund_canon_1} and \eqref{fundamental_canon_diff_forms}, which is
due to the different statistics of $\pa_\mu$ and $d$.}
\begin{align}\label{corner_EoM}
sC^\mu_\C-I_{V_\C}Z^\mu_{\rm cl}&=(-1)^{\epsilon(E_{\phi^i})}\pa_\nu\bigg[
E_{\phi^i}\left(c^B_1c^A_1\Big(\frac{\pa f^i_B}{\pa\pa_\nu\phi^j}f^{\mu j}_A
-M^{\mu C}_{BA}f^{\nu i}_C\Big)-2c^P_2M^{\mu B}_Pf^{\nu i}_B\right)\bigg]. 
\end{align}
In this calculation, we used \eqref{27}, \eqref{28}, \eqref{211}, \eqref{212}
and the Noether identity $E_{\phi^i}f^i_A-\pa_\nu(E_{\phi^i}f^{\nu i}_C)=0$.
The result \eqref{corner_EoM} was expected because from the definitions of
$C_\C$ and $Z$ in Appendix~\ref{Annexe_A}, one has
\begin{equation}\label{def_I}
d\big(sC_\C+I_{V_\C}Z\big)=0\quad\Longrightarrow\quad
sC_\C+I_{V_\C}Z\equiv dI.
\end{equation}
Therefore the classical part $I^{\mu\nu}_{\rm cl}$ of $I^{\mu\nu}$ is given
by the corner term \eqref{corner_EoM}. 
Notice that it has the exact same structure as the $F_{\phi^i}$ dependent
part of the gauge charges \eqref{explict_gauge_dependence}.  
We can finally use the definition of the equations of motion
\eqref{theta_gauge}
\begin{equation}
0\ \hat{=}\ \tilde{E}_{\phi^i}=E_{\phi^i}+(\text{terms\ proportional\ to }
\bar{c}_I\text{\ or\ }b_I)
\end{equation}
and the boundary conditions \eqref{bc_antighosts} to obtain 
\begin{equation}
\int_\varSigma (d^{D{-}1}x)_{\mu} \,\Big(sC^\mu_\C-I_{V_\C}Z^\mu_{\rm cl}\Big)
=\int_{\pa\varSigma} (d^{D{-}2}x)_{\mu\nu} \ I^{\mu\nu}_{\rm cl}\ \hat{=}\ 0.
\end{equation}
The same reasoning applies to the corner term $d i_\xi C_\C$ of
\eqref{computation_I_V_I_V_o}.  

Gathering everything, we find  
\begin{equation}\label{faithful_rep}
\boxed{{\{Q_\C,Q_\C\}}_L\ \hat{=}\ Q_{s\C}}\  
\end{equation}
for the bracket defined by \eqref{charge_bracket_local}.
One must appreciate that this bracket is constructed out of the gauge
dependent symplectic two-form $\tilde{\O}$ and leads, for any gauge
fixing functional~$\Ps$, to a faithful representation of the asymptotic
symmetry algebra~\eqref{asg_structure}.
This is a major result of this paper. 

The last property to check is the invariance of \eqref{faithful_rep} under
boundary shifts $L'=L+dl$.
The gauge dependent part of $\tilde{\o}$ is invariant under this shift but
the classical quantities transform as 
\begin{align}
\label{shift_transformations_1}
E'&=E, &C_\C'&=C_\C, &Z'_{\rm cl}&=Z_{\rm cl},                      \nn \\
\theta'&=\theta-\delta l+d\nu,&K'_\C&=K_\C-sl,
&q'_\C&=q_\C+I_{V_\C}\nu+i_\xi l,                                   \nn \\
a'_\C&=a_\C-\varDelta_\C l,&A'_\C&=A_\C-\varDelta_\C\nu,
&\kappa'&=\kappa+i_\phi l.                                           
\end{align}
By using these relations, one can check that 
\begin{align}
\label{shift_transformations_2}
\demi I_{V_\C} I_{V_\C} \tilde{\o}' &= \demi I_{V_\C} I_{V_\C} \tilde{\o}  - d I_{V_\C} s \nu , & d I_{V_\C} A'_\C &= d I_{V_\C} A_\C - d I_{V_\C} \varDelta_\C\nu ,
\nn \\
d \Lie_\xi q'_\C &=  d \Lie_\xi  q_\C + d  I_{V_\C} \Lie_\xi  \nu + d i_\xi d i_\xi l , & d I_{V_\C} q'_{\delta \C} &= d  q_{s \C} + d I_{V_\C} I_{V_{\delta \C}} \nu + d i_{s \xi} l ,
\nn \\
\demi di_\xi i_\xi L' &= \demi d i_\xi i_\xi L + \demi d i_\xi i_\xi dl  ,  &  d \kappa' &=  d \kappa + d i_\phi l  ,
\end{align}
with $di_\phi l=di_{s\xi}l+\demi di_\xi i_\xi dl-di_\xi d i_\xi l$ by the
definition \eqref{hat_s_2}.
For our bracket \eqref{charge_bracket_local}, we thus obtain
\begin{equation}
{\{Q'_\C,Q'_\C\}}_{L'}-Q'_{s\C}={\{Q_\C,Q_\C\}}_L-Q_{s\C}
\end{equation}
and therefore \eqref{boundary_shift_independence} is satisfied.

\subsection{BRST cocycle for asymptotic symmetries}
\label{section_BRST_cocycle}

The tools introduced in the previous sections allow us to 
systematically define a ghost number two BRST cocycle associated with 
asymptotic symmetries.
In this construction,  only the classical symplectic structure $\o = \delta \theta$ is relevant.  Indeed,  any gauge dependent quantity would depend on antighosts or auxiliary fields and therefore could only lead to trivial cocycles.

We use \eqref{def_Y_form} and \eqref{fundamental_canon_diff_forms}
to write 
\begin{equation}
I_{V_\C}I_{V_\C}\o=s\Big(-C_\C+dq_\C+I_{V_\C}\theta+K_\C\Big) ,
\end{equation}
where $s=L_{V_\C}$ is the large BRST operator.
From there we get 
\begin{equation}\label{s_fund_canon}
sI_{V_\C}I_{V_\C}\o=0.
\end{equation}
Then, by making use of \eqref{def_I} and of the classical part of the
computation \eqref{computation_I_V_I_V_o}, the equation \eqref{s_fund_canon}
leads to the off-shell identity 
\begin{equation}\label{s_fund_canon_explicit}
0=\demi sI_{V_\C}I_{V_\C}\o=sd\Big(\Lie_\xi q_\C-I_{V_\C}A_\C
 +\kappa-\demi i_\xi i_\xi L-i_\xi C_\C-I_{\rm cl}-q_{s\C}\Big).
\end{equation}
This means that if we define the ghost number two and spacetime codimension-two quantity 
\begin{equation}\label{BRST_2-cocycle}
\Delta^2_{d-2}\equiv\Lie_\xi q_\C-I_{V_\C}A_\C
+\kappa-\demi i_\xi i_\xi L-i_\xi C_\C-I_{\rm cl}-q_{s \C},
\end{equation}
then from \eqref{s_fund_canon_explicit} and the algebraic Poincar\'e lemma,
we obtain the descent equations 
\begin{align}\label{descent_equations}
s\Delta^2_{d-2} &= d \Delta^3_{d-3}, \nn \\
s\Delta^3_{d-3} &= d \Delta^4_{d-4}, \nn \\
  &\hspace{0.215cm}\vdots            \nn \\
s\Delta^{d-1}_1&=d\Delta^d_0.
\end{align}
It follows that $\Delta^2_{d-2}$ is a ghost number two BRST cocycle associated
with the large BRST operator $s=L_{V_\C}$  and thus with asymptotic symmetries.  
This cocycle justifies and extends the  model dependent constructions
of~\cite{Barnich_BRST,Baulieu_Tom_BMS}.

Moreover, $\Delta^2_{d-2}$ is actually half the Barnich--Troessaert bracket
defined in \eqref{generic_Q,Q}, because if we use \eqref{bc_antighosts},
\eqref{charge_bracket_local} and \eqref{BRST_2-cocycle}, we find 
\begin{equation}
{\{Q_\C,Q_\C\}}_L\ \hat{=}\ \frac12{\{Q_\C,Q_\C\}}_{\rm BT}
 +Q_{s\C}+\int_{\pa\varSigma}\Delta^2_{d-2} ,
\end{equation}
so that the important on-shell relation \eqref{faithful_rep} leads to
\begin{equation}
\label{relation_cocycle_bracket}
{\{Q_\C,Q_\C\}}_{\rm BT}\ \hat{=}\ -2 \int_{\pa\varSigma}\Delta^2_{d-2}. 
\end{equation}
The fact that, on-shell,  the cocycle \eqref{BRST_2-cocycle} is proportional to the BRST
Barnich--Troessaert bracket justifies the observation of
\cite{Baulieu_Tom_BMS,Baulieu:2024oql} that a top cocycle $\Delta^1_{d-1}$,
which is responsible for the perturbative corrections of
\eqref{charge_ward_identity} and which generates \eqref{descent_equations},
could sometimes be the corner charge $Q_\C$ itself.
A general construction of this $\Delta^1_{d-1}$ is
yet to be found.

\section{Application: Abelian 2-form coupled to Chern--Simons}
\label{Section_examples}

An interesting application of Noether's 1.5th theorem \eqref{Noether_1.5} is
for the gauge fixed theory of an abelian $2$-form $B_2$ coupled to
Chern--Simons in four-dimensions.
The Lagrangian is 
\begin{equation}\label{L_G3G3}
L=\demi G_3\star G_3
\end{equation}
with 
\begin{equation}
G_3=d B_2+\Tr\left(AdA+\frac23A^3\right). 
\end{equation}
The mixed abelian and non-abelian Chapline--Manton gauge symmetry of the
Lagrangian~\eqref{L_G3G3} is governed by two infinitesimal parameters,
a Lie algebra valued scalar $\eps$ and a real vector~$\eps_\mu$.
One defines $\eta\equiv(\eps,\epsilon_\mu)$ and one has  
\begin{align}\label{gaugeB_2}
\delta_\eta A_\mu&=D^A_\mu\eps ,                               \nn\\  
\delta_\eta B_{\mu\nu}&=\pa_{\{\mu}\eps_{\nu\}}+\Tr(\eps\pa_{\{\mu}A_{\nu\}}).
\end{align}
One can check that $\delta_\eta G_3=0$ and thus $\delta_\eta L=0$.
This symmetry is closed off-shell, namely
\begin{equation}
[\delta_\eta,\delta_{\eta'}]=\delta_{\eta''}\quad\text{with}\quad
 \eta''=\Big([\eps,\eps'],  \Tr(A_\mu[\eps,\eps']\Big).
\end{equation}

The theory defined by \eqref{L_G3G3} is thus a rank-$1$ BV system, which is
relevant in the framework of this paper for at least three reasons. 
First, it has non-trivial asymptotic symmetries associated with the gauge
symmetry of the $2$-form, which provide a symmetry interpretation for scalar
soft theorems \cite{Campiglia:2018see,Francia:2018jtb}.   
Second, its gauge fixing requires introducing ghosts of ghosts to the field
space, which justifies the inclusion of the fields $c^P_2$ in our general
derivations of Section~\ref{Section_BRST_Noether_current}.
Third, the curvature $G_3$ plays a fundamental role in the Green--Schwartz
anomaly compensation mechanism \cite{Green:1984sg,Baulieu:1985mb}. 
Computing the BRST Noether charge in this gauge fixed theory serves as a
relevant test to illustrate the power of our general formula
\eqref{Noether_1.5}.

The BRST field space is composed of an abelian $2$-form $B_{\mu\nu}$, its
associated ghost $\psi_\mu$ with a ghost for this ghost $\phi$, a non-abelian
$1$-form $A^a_\mu$ and its associated ghosts $c^a$ valued in a Lie algebra
$\mathfrak{g}$.
The nilpotent BRST transformations \eqref{BRST_trans} associated with the
gauge symmetry \eqref{gaugeB_2} that leave the Lagrangian \eqref{L_G3G3}
invariant are given by 
\begin{align}\label{s_B_2}
s A^a_\mu &= \pa_\mu c^a + f^a_{\ bc} A^b_\mu c^c ,
\nn \\
s c^a &= - \demi  f^a_{\ bc}  c^b c^c ,
\nn \\
s B_{\mu\nu} &= \pa_{\{ \mu} \psi_{\nu\} } + c^a \pa_{\{ \mu } A^a_{\nu \}} ,
\nn \\
s \psi_\mu &= \pa_\mu \phi - \demi A^a_\mu [c,c]^a , 
\nn \\
s \phi &= \frac{1}{6} c^a [c,c]^a ,
\end{align}
where $[c,c]^a=f^a_{\ bc}c^bc^c$ with the structure constants $f_{abc}$ of
$\mathfrak{g}$. 
These transformations naturally derive from the horizontality conditions on
$F=dA+A^2$ and $G_3$ \cite{Baulieu:1983bk} 
\begin{align}\label{geoB2}
\hat{F}&\equiv(d+s)(A+c)+(A+c)^2=F,                               \nn \\
\hat{G_3}&\equiv(d+s)(B_2+\psi^1_1+\phi^2_0)+\Tr\left((A + c )\hat{F}
 - \frac13(A+c)^3\right)=G_3,
\end{align}
which illustrate the geometrical significance of the gauge symmetry
\eqref{gaugeB_2}. 
The nilpotency is ensured by the Bianchi identities
\begin{align}
\hat{D}\hat{F}=0\quad\quad\text{and}\quad\quad
\hat{d}\hat{G}_3=\hat{F}\hat{F}=FF.
\end{align}
In particular, notice that the invariance of \eqref{L_G3G3} under
\eqref{s_B_2} is obvious because the ghost number one component of
$\hat{d}G_3=FF$ leads to $sG_3=0$.

We can now determine the functions parametrizing \eqref{BRST_trans} for the
specific case of \eqref{s_B_2}.
Since the fields entering \eqref{s_B_2} have certain indices in common and
sometimes have more than one index, the abstract indices $i$, $A$ and $P$ in 
\eqref{BRST_trans} will be replaced by the field itself.
For instance, if we compare 
\begin{align}
sA^a_\mu&=\pa_\mu c^a+f^a_{\ bc}A^b_\mu c^c\quad\quad\text{with}\quad\quad 
s\phi^i=f^i_Ac^A_1+f^{\mu i}_A\pa_\mu c^A_1,
\end{align}
we write $\phi^{A^a_\mu}=A^a_\mu$, $c^{c^a}_1=c^a$,
$c^{\psi_\rho}_1=\psi_\rho$ and thus we deduce  
\begin{align}\label{f_A_G3}
f^{A^a_\mu}_{c^c} &= f^a_{\ bc}A^b_\mu,&
f^{\nu A^a_\mu}_{c^b}&=\delta^\nu_\mu\delta^a_b,                 \nn\\
f^{A^a_\mu}_{\psi_\rho}&=0,&f^{\nu A^a_\mu}_{\psi_\rho}&=0.
\end{align}
We can do that for each transformations in \eqref{s_B_2}.
If we write only the non-vanishing functions, we have 
\begin{align}\label{f_B_G3}
f^{B_{\mu\nu}}_{c^a}&=\pa_{\{\mu}A^a_{\nu\}},
&f^{\rho B_{\mu\nu}}_{\psi_\gamma}
&=\delta^{\{\rho}_\mu\delta^{\gamma\}}_\nu,                            \nn \\
\gamma^{c^a}_{\ c^bc^c}&=-\demi f^a_{\ bc},
&\beta^\phi_{\ c^ac^bc^c}&=\frac16f^a_{\ bc},                          \nn \\
M^{\nu\psi_\mu}_\phi&=\delta^\nu_\mu,
&\gamma^{\psi_\mu}_{\ c^bc^c}&=-\demi f^a_{\ bc}A^a_\mu.
\end{align}

The next step is gauge fixing.
We must fix the gauge invariance associated with $c^a$, $\psi_\mu$ and $\phi$. 
A judicious choice of gauge for perturbation theory is for instance
\begin{align}\label{gauge_choice_B2}
\pa_\mu A^\mu_a+\demi b_a=0,\quad\quad
\pa^\nu B_{\mu\nu}+\pa_\mu L+\demi b_\mu=0,\quad\quad\pa^\mu\psi_\mu=0,
\end{align}
where we have introduced trivial BRST doublets $(\bar{c}_a^{-1} , b_a^0)$,
$(\bar{\psi}_\mu^{-1},b_\mu^0)$, $(\bar{\phi}^{-2},\bar{\eta}^{-1})$ and
$(L^0,\eta^1)$.
The upper index is the ghost number.
These doublets are captured by the last two lines of the parametrization
\eqref{BRST_trans}.
They allow us to impose the gauge condition \eqref{gauge_choice_B2} by
considering
\begin{equation}\label{L_GF_G3}
L_{\rm GF}=\demi G_3\star G_3
+s\left(\bar{c}^a\Big(\pa_\mu A^\mu_a+\demi b_a\Big)
+\bar{\psi}^\mu\Big(\pa^\nu B_{\mu\nu}+\pa_\mu L+\demi b_\mu\Big)
+\bar\phi\pa^\mu\psi_\mu\right).
\end{equation}

The only things we need in order to apply the BRST Noether theorem
\eqref{Noether_1.5} to this gauge fixed Lagrangian are the expressions of
$F_{\Ph^I}$ and $\Xi^\mu$ as defined in \eqref{fundam_identities} and the
expression of the classical Noether charge $q^{\mu\nu}_{\text{cl},c^A_1}$.
The classical charge is found by applying Noether's second theorem
\eqref{classical_J} to the ungauged Lagrangian \eqref{L_G3G3}.
In terms of spacetime differential forms, it is given by
\begin{align}
q_{\text{cl},c^A_1}=\Big(\psi^1_1+\Tr(cA)\Big)\star G_3. 
\end{align}
Then, by acting on the gauge fixing functional of \eqref{L_GF_G3} with
$\delta$, we find 
\begin{align}\label{Fs_G3}
F_{A^a_\mu}&=\pa^\mu\bar{c}_a,\quad\quad
F_{B_{\mu\nu}}=\pa^\nu\bar{\psi}^\mu,\quad\quad
F_{\psi_\mu}=-\pa^\mu\bar{\phi},\quad\quad
F_L=\pa_\mu\bar{\psi}^\mu,                                            \nn \\
\Xi^\mu&=-\bar{c}^a\delta A^\mu_a-\bar{\psi}^\mu\delta L \bar{\psi}^\nu
 \delta B_{\rho \nu}\eta^{\rho \mu}+\bar{\phi}\delta\psi^\mu.  
\end{align}
The expressions of $F_{\bar{c}_I}$ and $F_{b_I}$ are useless for our purpose
because these quantities don't appear in the BRST Noether current
\eqref{Noether_1.5}.
Notice however than $F_L$ does contribute because $L$ is a ``$\bar{c}_I$"
with ghost number zero, which means that the BRST transformation $sL$ enters
the parametrization \eqref{BRST_trans} through $s\phi^i$.
We thus have $f^L_\eta=1$. 

Finally, by plugging \eqref{f_A_G3}, \eqref{f_B_G3} and \eqref{Fs_G3} into
our general result \eqref{explict_gauge_dependence}, we get
\begin{multline}
J^\mu_{\rm BRST}
=\pa_\nu\Big(\big(\psi^1_1+\Tr(cA)\big)\star G_3\Big)^{\mu\nu}
 -2\pa_\nu\Big(\phi\pa^{\{\nu}\bar{\psi}^{\mu \}}\Big)               \\
\quad+s\Big(\bar{c}^asA^\mu_a+\bar{\psi}^\mu\eta+\bar{\psi}^\nu
 (sB_{\rho \nu})\eta^{\rho \mu}-\bar{\phi}s\psi^\mu+c^a\pa^\mu\bar{c}_a
 +\psi_\nu\pa^{\{\mu}\bar{\psi}^{\nu \}}+2\phi\pa^\mu\bar{\phi}\Big).
\end{multline}
In the free theory \eqref{L_G3G3} without the coupling to Chern--Simons,
the Ward identity \eqref{charge_ward_identity} $[Q,\mathcal{S}]=0$ for the
classical Noether charge $Q=\int\psi^1_1\star G_3$ would lead us to the scalar
soft theorem for the scalar field $\vp$ dual to the abelian $2$-form through
$\star d\vp=dB_2$ \cite{Campiglia:2018see,Francia:2018jtb}.
Here we have shown that a specific gauge fixing of this theory can lead to
non-trivial gauge dependent Noether charges
$Q_\phi=\int\phi\star(d\bar{\psi})$.
So without the general BRST construction of the Ward identity
$[Q,\mathcal{S}]=0$ discussed in Section~\ref{Holographic_WI} that ensures its
gauge independence, one could have naively thought that the charge $Q_\phi$
for the ghost of ghost must also commute with the $\mathcal{S}$-matrix,
therefore making the physical consequences of asymptotic symmetries gauge
dependent. 
This example thus confirms the need for a BRST construction of the holographic
Ward identity \eqref{charge_ward_identity}.
Notice that such gauge dependent charges also appear for asymptotically flat
gravity in the de~Donder gauge.
Their explicit expression was given at the end of Section~\ref{section_1.5}.

\section{Towards a generalization for rank-2 BV systems}
\label{Section_extension_rank-$2$}

So far we have only focused on rank-$1$ BV theories.
The symmetries of such theories are captured by a BRST operator which is
nilpotent off-shell and maps the fields on local field functionals with no
mixing with the antifields. 
This fact is reflected by a linear dependence on the antifields in the BV
action.  

However not all theories are of this type, as for instance some supergravity
theories without auxiliary fields whose infinitesimal gauge transformations
only close on-shell.
They appear as not being under the control of a genuinely field dependent BRST
operator which is nilpotent off-shell and they have a quadratic dependence on
the antifields in the BV action.
We thus say that such BV theories are of rank~2.
It is believed that all supergravities are at most BV systems of
rank~2.\footnote{To the best of our knowledge, we are not aware of any
physical theory with rank higher than two.}

Since these theories of interest have gauge symmetries, there are no reasons
they don't exhibit asymptotic symmetries \cite{Rejzner:2020xid}.
We should thus be able to study their physical effects through the holographic
Ward identities \eqref{charge_ward_identity}.   
Given that this Ward identity is derived from Noether's 1.5th theorem
\eqref{Noether_1.5}, we may question the validity of this theorem for  
rank-$2$ BV theories. 
Here we will only test the validity of the 1.5 Noether theorem for the modestly
simple rank-$2$ BV system, that is Yang-Mills theory in the Feynman-'t~Hooft
gauge without the $b$ field.  

We work in an $n$-dimensional Minkowski spacetime $M$ and start with the
usual rank-$1$ BV action of Yang--Mills theory, that is 
\begin{equation}\label{BV_action_YM_1}
L_{\rm BV}=-\frac14 F_{\mu\nu}F^{\mu\nu}-s\Ph^I{}^*\Ph_I,
\end{equation}
where we have introduced antifields ${}^*\Ph_I$ for each fields
$\Ph^I=\{A^a_\mu,c^a,\bar{c}_a,b_a\}$.  
If $p$ is the spacetime form degree and $g$ the ghost number, the grading of
an antifield ${}^*\Ph$ is given by
\begin{equation}
{}^*(\Ph^g_p)=({}^*\Ph)^{-g-1}_{n-p}.
\end{equation}
The antifields can be interpreted as sources for the nilpotent BRST
transformations of the fields
\begin{align}
sA^a_\mu &=D_\mu c^a=\pa_\mu c^a+[A_\mu,c]^a,&
s\bar{c}_a &=b_a,  \nn \\
sc^a &=-\demi[c,c]^a, & sb_a &= 0 , 
\end{align}
while their nilpotent BRST transformations are defined as 
\begin{equation}
s{}^*\Ph_I=\frac{\delta^L L_{\rm BV}}{\delta\Ph^I}. 
\end{equation}

To reach the Feynman-'t~Hooft gauge, we perform the following anticanonical
transformation 
\begin{align}
\left\{ 
\begin{aligned}
\Ph^I &=\tilde{\Ph}^I    \\
{}^*\Ph_I &={}^*\tilde{\Ph}_I+\frac{\delta^L\Ps}{\delta\Ph^I}  
\end{aligned}\right.
\end{align} 
for the gauge fixing functional
$\Ps=\bar{c}_a\big(\pa^\mu A^a_\mu+\demi b^a\big)$.  
Under this transformation, the BV action \eqref{BV_action_YM_1} becomes
\begin{equation}\label{L_BV_shifted1}
L_{\rm BV}=-\frac14 F_{\mu\nu}F^{\mu\nu}-s\Ph^I{}^*\tilde{\Ph}_I-s\Ps. 
\end{equation}
The ${}^*\tilde{\Ph}_I$'s can now be seen as background fields.
Setting them to zero fixes the gauge as in~\eqref{L_GF}.  
What will be useful is that \eqref{L_BV_shifted1} with the background fields
still satisfies the classical master equation
\begin{equation}\label{classical_master_equation}
s_\Ps S_{\rm BV}=2 \int_M d^n x \,    \frac{\delta^L S_{\rm BV}}{\delta \Ph^I(x)} \frac{\delta^R S_{\rm BV}}{\delta {}^*\tilde{\Ph}_I(x)} =\int_M dK_{\rm BV}
\end{equation}
for the shifted nilpotent BRST operator
\begin{align}\label{shifted_s_general}
s_\Ps\Ph^I &= s \Ph^I, \nn \\
s_\Ps{}^*\tilde{\Ph}_I
&=\frac{\delta^L(L_{\rm cl}-s\Ph^I{}^*\tilde{\Ph}_I-s\Ps)}{\delta\Ph^I}.
\end{align}

The original motivation was to calculate the Noether current of a Lagrangian
with a quadratic dependence in the antifields.
To get there, we integrate out the $b^a$ field in \eqref{L_BV_shifted1} and
obtain 
\begin{equation}\label{L_BV_shifted2}
L_{\rm BV}=-\frac14 F_{\mu\nu}F^{\mu\nu}-D_\mu c^a({}^*\tilde{A})^\mu_a
 +\demi[c,c]^a({}^*\tilde{c})_a+\bar{c}_a\pa_\mu D^\mu c^a-\demi
 \Big(\pa^\mu A_\mu^a+({}^*\tilde{\bar{c}})^a\Big)^2. 
\end{equation}
This Lagrangian is indeed quadratic in $({}^*\tilde{\bar{c}})^a$.
Its BRST symmetry is governed by the shifted operator
\eqref{shifted_s_general}, which takes the form
\begin{align}\label{s_shifted_YM}
s A^a_\mu &= D_\mu c^a   ,  &  s {}^*\tilde{A}^\mu_a &= D_\nu F^{\nu \mu}_a  + [\pa^\mu \bar{c} , c]_a - [{}^*\tilde{A}^\mu,c]_a + \pa^\mu \pa_\nu A^\nu_a + \pa^\mu {}^*  \tilde{\bar{c}}_a ,  
\nn \\
 s c^a &= - \demi [c,c]^a  ,   &    s {}^*\tilde{c}_a &= D^\mu \pa_\mu \bar{c}_a + [ {}^*\tilde{c} ,c]_a - D_\mu {}^*\tilde{A}^\mu_a  , 
\nn \\
 s \bar{c}^a &= \pa^\mu A_\mu^a + {}^*  \tilde{\bar{c}}^a , & s  {}^*  \tilde{\bar{c}}^a &= -\pa_\mu D^\mu c^a  .
\end{align}
The fact that we are dealing with a rank-$2$ BV system now becomes obvious
because if one fixes the gauge by setting the background fields to zero,
one gets $s^2\bar{c}^a=\pa^\mu D_\mu c^a\ne0$ off-shell.
The nilpotency is restored on-shell by the equation of motion of $\bar{c}_a$.  

Since the quantum field theory must explicitly depend on the background
fields for the master equation \eqref{classical_master_equation} to be valid,
we must compute the Noether current associated with the BRST symmetry
\eqref{s_shifted_YM} of the rank-$2$ BV Lagrangian \eqref{L_BV_shifted2}.
We have 
\begin{align}
\delta L_{\rm BV}&=\tilde{E}_{\Ph^I}\delta\Ph^I
 +(-1)^{\epsilon(\Ph^I)}s\Ph^I\delta({}^*\tilde{\Ph}_I)
 +\pa_\mu\tilde{\theta}^\mu_{\rm BV},                                 \nn \\
sL_{\rm BV}&=\pa_\mu K^\mu_{\rm BV}.
\end{align}
This provides
\begin{align}\label{theta_K_BV_YM}
\tilde{\theta}^\mu_{\rm BV}
&=-\delta A_\nu^a F^{\mu\nu}_a-\delta A^\mu_a
 \big([\bar{c},c]^a+{}^*\tilde{\bar{c}}^a+\pa^\nu A_\nu^a\big)
 -\delta c^a{}^*\tilde{A}^\mu_a-\bar{c}_aD^\mu(\delta c^a)
 +\pa^\mu\bar{c}_a\delta c^a,                                        \nn \\
K^\mu_{\rm BV}&=c^aD_\nu F^{\nu\mu}_a+\demi[c,c]^a
 \big(\pa^\mu\bar{c}_a-{}^*\tilde{A}^\mu_a\big)+D^\mu c^a
 \big(\pa_\nu A^\nu_a+{}^*\tilde{\bar{c}}_a\big). 
\end{align}
We can now compute the Noether current by defining the vector field $V$ on
the field space generating the transformations \eqref{s_shifted_YM} and
considering
\begin{equation}\label{BV_current_generic}
J^\mu_{\rm BV}=I_V\tilde{\theta}^\mu_{\rm BV}-K^\mu_{\rm BV}
=-C^\mu_{\rm BV}+\pa_\nu q^{\mu\nu}_{\rm BV},
\end{equation}
where $\pa_\mu C^\mu_{\rm BV}=I_V\Big(\tilde{E}_{\Ph^I}\delta \Ph^I
+(-1)^{\epsilon(\Ph^I)}s\Ph^I\delta({}^*\tilde{\Ph}_I)\Big)$ and
$q^{\mu\nu}_{\rm BV}$ has yet to be determined.
We find 
\begin{align}\label{s-exact_BV}
C^\mu_{\rm BV}=2K^\mu_{\rm BV}\ \hat{=}\
2s\Big(c^a(\pa^\mu\bar{c}_a-{}^*\tilde{A}^\mu_a)+\bar{c}_aD^\mu c^a\Big),
\end{align}
where we have used the field equations of motion that are given by the BRST
transformations of the antifields in \eqref{s_shifted_YM}.  
Then by using \eqref{theta_K_BV_YM} and plugging the off-shell expression of
$C^\mu$ in \eqref{BV_current_generic}, we get
\begin{equation}
q^{\mu\nu}_{\rm BV}=-c^aF^{\mu\nu}_a. 
\end{equation} 
The BRST Noether 1.5th theorem \eqref{Noether_1.5} is thus valid for this
simple rank-$2$ BV system. 

Whether or not this simple example really tells us something non-trivial
about rank-$2$ BV systems is unclear.
In fact, the structure of the $s$-exact term \eqref{s-exact_BV} resembles that
of the rank-$1$ theory, which consist in keeping the $b^a$ field and imposing
the Feynman-'t~Hooft gauge in the usual way \eqref{L_GF}.
By doing so, one obtains \cite{Baulieu:2024oql}
\begin{equation}
C^\mu\ \hat{=}\ s(c^a\pa^\mu\bar{c}_a+b^aA^\mu_a).
\end{equation}
It has indeed the same structure as \eqref{s-exact_BV} if one sets the
antifields to zero and reintroduce the $b^a$ field because in this case
$s(\bar{c}_aD^\mu c^a)=b_aD^\mu c^a=s(b^aA^\mu_a)$.

Proving the BRST Noether 1.5th theorem for every rank-$2$ BV theories will
be the subject of a future publication. 

\begin{center}
\textbf{Acknowledgments}
\end{center}
\indent It is a pleasure to thank Glenn Barnich, Marc Bellon and
Luca Ciambelli for insightful discussions.
The work of S.W. is supported in part by the NSTC grants
Nos.~112-2115-M-007-006 and 114-2115-M-007-003.

\appendix

\section{Constraints on the BRST transformations}
\label{Annexe_constraints}

By making use of \eqref{s2cphi}, \eqref{s2c1} and \eqref{s2c2}, the
parametrizing functions of the BRST transformations \eqref{BRST_trans} must
satisfy the following constraints:
{\allowdisplaybreaks
\begin{align}
\label{26}
&\bullet \ f^i_A \gamma^A_P + f^{\mu i}_A \pa_\mu \gamma^A_P = 0  
\\ \label{27}
&\bullet \ f^{\mu i}_A  M^{\nu A}_P \ \text{is antisymmetric in} \ (\mu,\nu) 
\\  \label{28}
&\bullet \ f^i_A M^{\mu A}_P  + f^{\mu i }_A \gamma^A_P + f^{\nu i}_A \pa_\nu M^{\mu A}_P =0 
\\  \label{29}
&\bullet \  \frac{\pa f^i_A}{\pa \phi^j	} f^j_B + \frac{\pa f^i_A}{\pa \pu \phi^j} \pu f^j_B  + f^i_C \gamma^C_{\ BA} + \fiu_C \pu \gamma^C_{\ BA} \ \text{is symmetric in} \ (A,B) 
\\ \label{210}
&\bullet  \ \frac{\pa f^i_A}{\pa \phi^j	} f^{\mu j}_B + \frac{\pa f^i_A}{\pa \pu \phi^j} f^j_B + \frac{\pa f^i_A}{\pa \pv \phi^j} \pv f^{\mu j}_B - f^i_C M^{\mu C}_{AB} - \frac{\pa \fiu_B}{\pa \phi^j} f^j_A + 2 \fiu_C \gamma^C_{\ BA} - f^{\nu i}_C \pv M^{\mu C}_{AB} = 0 
\\ \label{211}
&\bullet \ \frac{\pa f^i_A}{\pa \pu \phi^j} f^{\nu j}_B - \fiu_C M^{\nu C}_{AB} \ \text{is antisymmetric in} \ (\mu,\nu) 
\\ \label{212}
&\bullet \ \frac{\pa \fiu_A}{\pa \phi^j} f^{\nu j}_B + f^{\nu i}_C M^{\mu C}_{BA} \ \text{is symmetric in} \ \big( (\mu,A) , (\nu, B) \big)
\\ \label{213}
&\bullet \   \frac{\pa \gamma^A_{\ P}}{\pa \phi^j} f^j_D + \gamma^A_{\ Q} \varGamma^Q_{\ PD} + M^{\mu A}_Q \pa_\mu \varGamma^Q_{\ PD} + \gamma^A_{\ \{CD\}} \gamma^C_{\ P} - M^{\mu A}_{DC} \pa_\mu \gamma^C_{\ P}  = 0  
\\
\label{214}
&\bullet \ \frac{\gamma^A_{\ P}}{\pa \phi^j} f^{\mu j }_D  + M^{\mu A}_Q \varGamma^Q_{\ PD} + M^{\mu A}_{CD} \gamma^C_{\ P} =0 
\\ \label{215}
&\bullet \ M^{\mu A}_Q \varGamma^{\nu Q}_{DP} - M^{\mu A}_{DQ} M^{\nu Q}_P \ \text{is antisymmetric in} \ (\mu,\nu) 
\\  \nn
&\bullet \ \frac{\pa M^{\mu A}_P}{\pa \phi^j} f^j_D + \gamma^A_{\ Q} \varGamma^{\mu Q}_{DP} + M^{\nu A}_Q \pa_\nu \varGamma^{\mu Q}_{DP} + M^{\mu A}_Q \varGamma^Q_{PD} + \gamma^A_{\ \{CD \} } M^{\mu C}_P  - M^{\mu A}_{DC} \gamma^C_{\ P} 
\\  \label{216}
&\quad - M^{\nu A}_{DC} \pa_\nu M^{\mu C}_P = 0 
\\  \label{217}
&\bullet \  \frac{\pa M^{\mu A}_P}{\pa \phi^j} f^{\nu j}_D + M^{\nu A}_Q \varGamma^{\mu Q}_{DP} + M^{\nu A}_{CD} M^{\mu C}_P = 0 
\\  \label{218}
&\bullet \    \gamma^A_{\ P} \beta^P_{\ DBC} + M^{\mu A}_P \pa_\mu \beta^P_{\ DBC} + \frac{\pa \gamma^A_{\ BC}}{\pa \phi^j} f^j_D +  \gamma^A_{\ \{EC\} } \gamma^E_{\ DB} + M^{\mu A}_{BE} \pu \gamma^E_{\ DC}  \nn \\ &\quad \  \text{is totally symmetric in} \ (D,B,C)   
\\ \label{219}
&\bullet \   \frac{\pa \gamma^A_{\ BC}}{\pa \phi^j} f^{\mu j}_D + \frac{\pa M^{\mu A}_{CD}}{\pa \phi^j} f^i_B -  \gamma^A_{\ \{EC\} } M^{\mu E}_{BD} +  \gamma^E_{\ \{DC\} } M^{\mu A}_{BE} +  \gamma^E_{\ BC} M^{\mu A}_{ED} - M^{\nu A}_{BE} \pa_\nu M^{\mu E}_{CD} \nn \\  &\quad + M^{\mu A}_E \beta^E_{\ \{DBC\} }   \ \text{is symmetric in} \ (B,C)  
\\ \label{220}
&\bullet \  M^{\mu A}_{B E} M^{\nu E}_{C D} \ \text{is symmetric in} \ (B,C) \ \text{and/or antisymmetric in} \ (\mu,\nu)    
\\ \label{221}
&\bullet \   M^{\mu A}_{ E C} M^{\nu E}_{D B} - M^{\nu A}_{D E} M^{\mu E}_{ B C} - \frac{\pa M^{\mu A}_{DC}}{\pa \phi^j} f^{\nu j}_B \ \text{is symmetric in} \ \big( (\mu,C) , (\nu, B) \big)   
\\ \label{222}
&\bullet \  \varGamma^P_{\ QB} \gamma^B_{\ R} \ \text{is antisymmetric in} \ (Q,R)  
\\ \label{223}
&\bullet \  \varGamma^P_{\ QB} M^{\nu B}_{R} + \varGamma^{\nu P}_{BR} \gamma^B_{\ Q} = 0   
\\ \label{224}
&\bullet \   \varGamma^{\mu P}_{BQ} M^{\nu B}_R \ \text{is antisymmetric in} \ \big( (\mu,Q) , (\nu, R) \big)  
\\ \nn
&\bullet \  - \frac{\pa \varGamma^P_{\ QC}}{\pa \phi^j} f^j_E  - \varGamma^P_{\ RC} \varGamma^R_{\ QE} - \varGamma^{\mu P}_{CR} \pa_\mu \varGamma^R_{\ QE} + \beta^P_{\ \{DCE\}} \gamma^D_{\ Q}  + \varGamma^P_{\ QD} \gamma^D_{\ CE} \ \text{is symmetric} 
\\ \label{225}
&\quad \ \text{in} \ (C,E)     
\\ \label{226}
&\bullet \    \varGamma^{\mu P}_{CQ} \varGamma^{\nu Q}_{DR} \ \text{is symmetric in} \ (C,D) \ \text{and/or antisymmetric in} \ (\mu,\nu) 
\\ \label{227}
&\bullet \  - \frac{\pa \varGamma^{\mu P}_{BQ}}{\pa \phi^j} f^{\nu j}_E  - \varGamma^{\nu P}_{BR} \varGamma^{\mu R}_{EQ} + \varGamma^{\mu P}_{DQ} M^{\nu D}_{BE} = 0   
\\ \label{228}
&\bullet \   - \frac{\pa \varGamma^P_{\ QC}}{\pa \phi^j} f^{\nu j }_E   - \varGamma^{\nu P}_{CR} \varGamma^R_{\ QE} + \varGamma^P_{\ QD} M^{\nu D}_{CE } = 0   
\\ \nn
&\bullet \  - \frac{\pa \varGamma^{\mu P}_{BQ}}{\pa \phi^j} f^j_E - \varGamma^P_{\ RB} \varGamma^{\mu R}_{EQ} - \varGamma^{\mu P}_{BR} \varGamma^R_{\ QE} -\varGamma^{\nu P}_{BR} \pa_\nu \varGamma^{\mu R}_{EQ}  + \beta^P_{\ \{DBE\}} M^{\mu D}_Q + \varGamma^{\mu P}_{DQ} \gamma^D_{\ BE}   \ \text{is}  
\\ \label{229}
&\quad \  \text{symmetric in} \ (B,E) 
\\ \nn
&\bullet \    - \frac{\pa \beta^P_{\ BCD}}{\pa \phi^j} f^j_E -\varGamma^P_{\ QB} \beta^Q_{\ CDE} - \varGamma^{\mu P}_{BQ} \pa_\mu \beta^Q_{\ CDE} + \beta^P_{\ \{ABC\}} \gamma^A_{\ DE} \  \text{is totally symmetric} 
\\ \label{230}
&\quad \ \text{in} \ (B,C,D,E)    
\\ \label{231}
&\bullet \   - \frac{\pa \beta^P_{\ BCD}}{\pa \phi^j} f^{\nu j}_E - \varGamma^{\nu P}_{BQ} \beta^Q_{\ \{ECD\}} + \beta^P_{\ \{ABC\}} M^{\nu A}_{DE}  \  \text{is totally symmetric in} \ (B,C,D)  .
\end{align} }

It is important to notice that these constraints are the least restrictive
ones.  
In fact, we have assumed that whenever a product of ghosts of the type
$c^A_1 c^B_1$ or $c^P_2 c^Q_2$ appears in \eqref{s2cphi} or \eqref{s2c2},
one has
\begin{align}
c^A_1 c^B_1 f_{AB} &= - c^B_1 c^A_1 f_{AB} = - c^A_1 c^B_1 f_{BA}, \nn \\
c^P_2 c^Q_2 g_{PQ} &=  c^Q_2 c^P_2 g_{PQ} =  c^P_2 c^Q_2 g_{QP},
\end{align}
for all tensors $f_{AB}$ and $g_{PQ}$.
This means that the various products of ghosts showing up in \eqref{s2cphi},
\eqref{s2c1} or \eqref{s2c2} involve only ghosts of the same type so that we
can exchange their indices $A,B,\dots$, or $P,Q,\dots$, at will.\footnote{By
the same type, we mean that they are associated with the same gauge (or gauge
for gauge) symmetry.}
This is the reason why we obtained some commutation or anticommutation
restrictions on the functions of the parametrization \eqref{BRST_trans}. 

The computations of the paper remain valid in more realistic cases where
different types of ghosts appear and mix, when the above constraints are more
restrictive and give vanishing properties rather than (anti)symmetric
conditions.

\section{Review of the BRST covariant phase space}\label{Annexe_A}

In this Appendix, we recall some basic properties of the BRST covariant phase
space introduced in \cite{Baulieu:2024oql}.

We consider a field theory on a $d$-dimensional spacetime $M$ with Lagrangian
$\tilde{L}$ given by \eqref{L_GF} and a BRST field space $\tilde{\mathcal{F}}$
containing the fields of \eqref{BRST_trans}.
We can then define trigraded \textit{local forms} $X^{g,q}_p$ on
$\tilde{\mathcal{F}}\times M$ by making use of the three compatible
differentials $\delta$, $d$ and $s$, where $\delta$ can be thought of as an
arbitrary variation in the field space, $d$ is the exterior derivative on the
spacetime and $s$ is the BRST differential \eqref{BRST_trans}.
They satisfy 
\begin{equation}\label{trigrading}
(\delta+d+s)^2=0.
\end{equation}
If $g$ is the ghost number, $q$ the field space form degree and $p$ the
spacetime form degree, then any local form $X^{g,q}_p$ has a total grading
\begin{equation}
g(X^{g,q}_p)=g+q+p
\end{equation}
that defines its statistics. 
For instance, the Lagrangian $\tilde{L}=\tilde{L}^{0,0}_d$ is a spacetime top
form whose grading is equal to $d$.
We also define the graded commutator between any local forms $X$ and $Y$ as 
\begin{equation}\label{graded_commutator1}
[X,Y]=XY-(-1)^{g(X)g(Y)}YX.
\end{equation}

In the BRST covariant phase space,  as explained at the beginning of  Section~\ref{section_1.5},  the BRST operator $s$ acting on the whole covariant phase space and
satisfying \eqref{trigrading} is defined~as
\begin{equation}\label{defining_s}
s\equiv L_V=I_V\delta-\delta I_V.
\end{equation}
Here $L_V$ is the field space Lie derivative along the specific ghost number one vector field $V\equiv V_{\text{\tiny BRST}}$ defined in
\eqref{V_BRST_main} and $I_V$ is the interior product in field space. 

Let us introduce some other objects of the BRST covariant phase space that
are relevant to this work. 
In Section~\ref{section_BRST_flux}, we deal with ghost fields $c^A_1$ (in
the notation of \eqref{BRST_trans}) that are also vector fields in spacetime.  
They are the ghosts $\xi$'s associated with reparametrization symmetry.   
This means that one can consider the Lie derivative $\mathcal{L}_\xi$ in
spacetime along this vector ghost.
Because of the odd statistics of $\xi$, this Lie derivative writes
\begin{equation}\label{lie_xi}
\Lie_\xi=i_\xi d-di_\xi,
\end{equation}
where $i_\xi$ is the interior product.
The spacetime Lie derivative \eqref{lie_xi} appears in the BRST
transformations \eqref{BRST_trans} of the fields when the theory under study
is invariant under diffeomorphism.  

In some theories, such as topological gravity or supergravity, one can also
encounter ghost number two fields $c^P_2$ that are vector fields $\phi$ in
spacetime.\footnote{In supergravity, one has
$\phi^\mu=-\demi i\bar{\chi}\gamma^\mu\chi$ where $\chi$ is the ghost for
local supersymmetry \cite{Baulieu:1985md}.}
They appear in the BRST transformation of the $\xi$'s as 
\begin{equation}
s\xi=\demi\Lie_\xi\xi+\phi. 
\end{equation}
The nilpotency of the BRST operator requires that $\phi$ satisfies
$\{\phi,\phi\}=0$ and that it transforms under BRST symmetry as
\begin{equation}
\label{s_phi}
s\phi=\Lie_\xi\phi=\{\xi,\phi\}.
\end{equation}

The trigraded commutators \eqref{graded_commutator1} of the BRST covariant
phase space that are relevant to us are listed in Table~\ref{table_2}.
We now use them to construct the conserved BRST Noether current and the BRST
fundamental canonical relation associated with the gauge fixed Lagrangian
\eqref{L_GF}.
\begin{table}[ht!] 
\centering
\begin{tcolorbox}[tab,tabularx={X||Y|Y|Y|Y|Y|Y|Y|Y},boxrule=0.9pt] 
$ \;\;[\,\cdot\,,\,\cdot\,] $ & $ \hfill d \hfill$     & $\hfill \delta \hfill $     & $\hfill s = L_V \hfill$   & $\hfill I_V \hfill$     &   $\hfill \Lie_\xi \hfill$    & $\hfill \Lie_\phi \hfill$   & $\hfill i_\xi \hfill$     & $\hfill i_\phi \hfill$   
 \\\hline\hline
$\hfill d \hfill$   &  $\hfill 0 \hfill$     &  $\hfill 0 \hfill$      & $\hfill 0 \hfill$   & $\hfill 0 \hfill$     & $\hfill 0 \hfill$      & $\hfill 0 \hfill$   & $\hfill  -\Lie_\xi \hfill $     & $\hfill  -\Lie_\phi\hfill $
 \\\hline
$\hfill \delta \hfill$ & $\hfill 0 \hfill$     & $\hfill 0 \hfill$     & $\hfill 0 \hfill$   & $\hfill  - L_V \hfill $     & $\hfill  \Lie_{\delta \xi} \hfill $      & $\hfill  \Lie_{\delta \phi} \hfill $   & $\hfill  i_{\delta \xi} \hfill $     & $\hfill  i_{\delta \phi} \hfill $ 
\\\hline
$\hfill s = L_V \hfill$  & $\hfill 0 \hfill$     & $\hfill 0 \hfill$    & $\hfill 0 \hfill$   & $\hfill 0 \hfill$    & $\hfill  \Lie_{s\xi} \hfill $      & $\hfill  \Lie_{\{\xi,\phi\}} \hfill $   & $\hfill  i_{s \xi}\hfill $     & $\hfill  i_{\{\xi,\phi\}} \hfill $
\\\hline
$\hfill I_V \hfill$  & $\hfill 0 \hfill$     & $\hfill  L_V \hfill $     & $\hfill 0 \hfill$   & $\hfill 0 \hfill$     & $\hfill 0 \hfill$      & $\hfill 0 \hfill$   & $\hfill 0 \hfill$     & $\hfill 0 \hfill$ 
\\ \hline
$\hfill \Lie_\xi \hfill $  & $\hfill 0 \hfill$     & $\hfill  \Lie_{\delta \xi} \hfill $    & $\hfill  \Lie_{s\xi} \hfill $   & $\hfill 0 \hfill$     & $\hfill  \Lie_{\{ \xi, \xi \}} \hfill $     & $\hfill  \Lie_{\{ \xi, \phi \}} \hfill $   & $\hfill  i_{\{\xi,\xi\}} \hfill  $     & $\hfill  i_{\{\xi,\phi\}} \hfill $
\\\hline
$\hfill \Lie_\phi \hfill$  & $\hfill 0 \hfill$     & $\hfill  -\Lie_{\delta \phi} \hfill $     & $\hfill  -\Lie_{\{\xi,\phi\}} \hfill $   & $\hfill 0 \hfill$     & $\hfill  \Lie_{\{\xi,\phi \}} \hfill $      & $\hfill  \Lie_{\{\phi,\phi \}} \hfill $   & $\hfill  -i_{\{\xi,\phi\}} \hfill $     & $\hfill 0 \hfill$
\\\hline
$\hfill i_\xi \hfill$  & $\hfill  \Lie_\xi \hfill $     & $\hfill  -i_{\delta \xi} \hfill $     & $\hfill  -i_{s\xi} \hfill $   & $\hfill 0 \hfill$     & $\hfill  - i_{\{\xi,\xi\}} \hfill $      & $\hfill  i_{\{\xi,\phi\}} \hfill $   & $\hfill 0 \hfill$     & $\hfill 0 \hfill$
\\\hline
$\hfill i_\phi \hfill$ & $\hfill  \Lie_\phi \hfill $     & $\hfill  i_{\delta\phi} \hfill $     & $\hfill  i_{\{\xi,\phi\}} \hfill $   & $\hfill 0 \hfill$     &  $\hfill  i_{\{\xi,\phi\}} \hfill $      & $\hfill 0 \hfill$   & $\hfill 0 \hfill$     & $\hfill 0 \hfill$
\end{tcolorbox}
\caption{\textit{Graded commutators in the BRST covariant phase space} \label{table_2} }
\end{table}

We start from the fundamental identities of the BRST covariant phase that
define the equations of motion of the gauge fixed Lagrangian \eqref{L_GF}
and its BRST invariance
\begin{align}\label{fund_identity_EoM}
\delta\tilde{L}&=\tilde{E}+d\tilde{\theta},    \\
\label{BRST_inv_L} s\tilde{L}&=dK.
\end{align}
Here we have defined $\tilde{E}\equiv\tilde{E}_{\Ph^I}\delta\Ph^I$. 
The BRST Noether current associated with the invariance \eqref{BRST_inv_L}
is obtained by contracting the identity \eqref{fund_identity_EoM} with $I_V$.
Using \eqref{defining_s} and \eqref{BRST_inv_L}, this leads to 
\begin{align}\label{I_VE_def_C}
dK&=I_V\tilde{E}+dI_V\tilde{\theta}\ \hat{=}\ dI_V\tilde{\theta}
\quad\Longrightarrow\quad d(I_V\tilde{\theta}-K)\ \hat{=}\ 0.
\end{align}
We thus define the conserved BRST Noether current as 
\begin{equation}\label{def_Noether_J}
J_{\rm BRST}\equiv I_V\tilde{\theta}-K. 
\end{equation}
One then uses \eqref{I_VE_def_C} to write $I_V \tilde{E} \equiv dC$ and the
algebraic Poincar\'e lemma on the spacetime to define the corner Noether
charges $q=q^{1,0}_{d-2}$ of the current \eqref{def_Noether_J} as 
\begin{equation}\label{def_Noether_J_dq}
J_{\rm BRST}=I_V\tilde{\theta}-K=-C+dq. 
\end{equation}
In the  ungauge fixed case, one finds $C=E_{\phi^i}f^{\mu i}_Ac^A_1$ for
the BRST symmetry \eqref{BRST_trans} and we get Noether's second theorem
\eqref{classical_J}. 

We now turn to the BRST fundamental canonical relation.
We want to determine the expression of $I_V \tilde{\o}$ for the local
symplectic two-form $\tilde{\o}\equiv\delta\tilde{\theta}$.
To do so, we first introduce the local form $Z=Z^{1,1}_{d-1}$ by using
\eqref{fund_identity_EoM}, \eqref{BRST_inv_L} and   consider 
\begin{equation}
d\delta K=s\delta\tilde{L}=s\tilde{E}-ds\tilde{\theta}
\quad\Longrightarrow\quad s\tilde{E}\equiv dZ.
\end{equation}
From there, we can also introduce $Y=Y^{1,1}_{d-2}$ by using the algebraic
Poincar\'e lemma 
\begin{equation}\label{def_Y_form}
d(s\tilde{\theta}+\delta K-Z)=0\quad\Longrightarrow\quad
s\tilde{\theta}+\delta K-Z\equiv dY.
\end{equation}
Finally, by using \eqref{def_Noether_J_dq}, \eqref{def_Y_form}, the definition
of $s$ in \eqref{defining_s} and that of the local symplectic two-form
$\tilde{\o}$, we get 
\begin{align}
dY+Z-\delta K=s\tilde\theta=I_V\tilde{\o}-\delta I_V \tilde{\theta}
=I_V\tilde{\o}-\delta\big(-C+dq+K\big).
\end{align} 
This provides the BRST fundamental canonical relation, that is 
\begin{equation}\label{fundamental_canon_diff_forms}
I_V\tilde{\o}=\delta\big(-C+dq\big)+Z+dY,
\end{equation} 
which proves \eqref{GF_fund_canon_1} and \eqref{def_Z_Y}.

\bibliography{biblio}
\bibliographystyle{JHEP}

\end{document}